\documentclass[twocolumn,twocolappendix,usenames,dvipsnames]{aastex63}
\usepackage{amsmath}

\shorttitle{Physical Drivers of Disk Spin-up}
\shortauthors{Semenov et al.}
\graphicspath{{./}{figures/}}

\usepackage{etoolbox}
\makeatletter
\patchcmd{\NAT@citex}
  {\@citea\NAT@hyper@{\NAT@nmfmt{\NAT@nm}\NAT@date}}
  {\@citea\NAT@nmfmt{\NAT@nm}\NAT@hyper@{\NAT@date}}
  {}% Do nothing if patch works
  {}% Do nothing if patch fails
\patchcmd{\NAT@citex}
  {\@citea\NAT@hyper@{%
     \NAT@nmfmt{\NAT@nm}%
     \hyper@natlinkbreak{\NAT@aysep\NAT@spacechar}{\@citeb\@extra@b@citeb}%
     \NAT@date}}
  {\@citea\NAT@nmfmt{\NAT@nm}%
   \NAT@aysep\NAT@spacechar%
   \NAT@hyper@{\NAT@date}}
  {}% Do nothing if patch works
  {}% Do nothing if patch fails
\patchcmd{\NAT@citex}
  {\@citea\NAT@hyper@{%
     \NAT@nmfmt{\NAT@nm}%
     \hyper@natlinkbreak{\NAT@spacechar\NAT@@open\if*#1*\else#1\NAT@spacechar\fi}%
       {\@citeb\@extra@b@citeb}%
     \NAT@date}}
  {\@citea\NAT@nmfmt{\NAT@nm}%
   \NAT@spacechar\NAT@@open\if*#1*\else#1\NAT@spacechar\fi%
   \NAT@hyper@{\NAT@date}}
  {}% Do nothing if patch works
  {}% Do nothing if patch fails
\makeatother

\def\Mvir{M_{\rm vir}}
\def\Rvir{R_{\rm vir}}
\def\Tvir{T_{\rm vir}}
\def\vvir{v_{\rm vir}}
\def\vrot{v_{\rm rot}}
\def\vc{v_{\rm c}}
\def\cs{c_{\rm s}}

\def\Rsfr{R_{\rm sfr}}
\def\Rcirc{R_{\rm circ}}
\def\Rvmax{R_{v_{\rm c}, {\rm max}}}

\def\jzjc{j_z/j_{\rm c}}
\def\Mach{\mathcal{M}}

\def\SFR{\dot{M}_{\star}}
\def\tspinup{t_{\rm spin\text{-}up}}

\def\pc{{\rm \;pc}}
\def\kpc{{\rm \;kpc}}

\def\kms{{\rm \;km\;s^{-1}}}
\def\Msunyr{{\rm \;M_\odot\;yr^{-1}}}
\def\Msun{{\rm \;M_\odot}}

\def\Gyr{{\rm \;Gyr}}
\def\cc{{\rm \;cm^{-3}}}
\def\K{{\rm \;K}}

\def\mp{m_{\rm p}}
\def\kboltz{k_{\rm b}}

\defcitealias{bk22}{BK22}
\defcitealias{semenov23a}{Paper~I}

\newcommand{\newtext}[1]{#1}

\begin{document}

\title{Formation of Galactic Disks II: the Physical Drivers of Disk Spin-up}

\author[0000-0002-6648-7136]{Vadim A. Semenov}
\altaffiliation{\href{mailto:vadim.semenov@cfa.harvard.edu}{vadim.semenov@cfa.harvard.edu} \\ NHFP Hubble Fellow}
\affiliation{Center for Astrophysics $|$ Harvard \& Smithsonian, 60 Garden St, Cambridge, MA 02138, USA}

\author[0000-0002-1590-8551]{Charlie Conroy}
\affiliation{Center for Astrophysics $|$ Harvard \& Smithsonian, 60 Garden St, Cambridge, MA 02138, USA}

\author[0000-0002-0572-8012]{Vedant Chandra}
\affiliation{Center for Astrophysics $|$ Harvard \& Smithsonian, 60 Garden St, Cambridge, MA 02138, USA}

\author[0000-0001-6950-1629]{Lars Hernquist}
\affiliation{Center for Astrophysics $|$ Harvard \& Smithsonian, 60 Garden St, Cambridge, MA 02138, USA}

\author[0000-0001-8421-5890]{Dylan Nelson}
\affiliation{Universitat Heidelberg, Zentrum f{\"u}r Astronomie, Institut f{\"u}r theoretische Astrophysik, Albert-Ueberle-Str. 2, D-69120 Heidelberg, Germany}

\begin{abstract}
Using a representative sample of Milky Way (MW)-like galaxies from the TNG50 cosmological simulation, we investigate physical processes driving the formation of galactic disks. A disk forms as a result of the interplay between inflow and outflow carrying angular momentum in and out of the galaxy.
Interestingly, the inflow and outflow have remarkably similar distributions of angular momentum, suggesting an exchange of angular momentum and/or outflow recycling, leading to continuous feeding of prealigned material from the corotating circumgalactic medium. 
We show that the disk formation in TNG50 is correlated with stellar bulge formation, in qualitative agreement with a recent theoretical model of disk formation facilitated by steep gravitational potentials. Disk formation is also correlated with the formation of a hot circumgalactic halo with \newtext{around half} of the inflow occurring at subsonic and transonic velocities \newtext{corresponding to Mach numbers of $\lesssim2$}. In the context of recent theoretical works connecting disk settling and hot halo formation, our results imply that the subsonic part of the inflow may settle into a disk while the remaining supersonic inflow will perturb this disk via the chaotic cold accretion. 
We find that disks tend to form when the host halos become more massive than $\sim (1\text{--}2) \times 10^{11} \Msun$, consistent with previous theoretical findings and observational estimates of the predisk protogalaxy remnant in the MW.
Our results do not prove that either corotating outflow recycling, gravitational potential steepening, or hot halo formation \emph{cause} disk formation, but they show that all these processes occur concurrently and may play an important role in disk growth.
\end{abstract}

%-----------------------------------------------------------------
%-----------------------------------------------------------------

%-----------------------------------------------------------------
%-----------------------------------------------------------------
\section{Introduction}
%-----------------------------------------------------------------

Most of the star-forming galaxies in the present-day universe are disks. Until recently, such disk galaxies were thought to form recently, at redshifts $z \lesssim 1\text{--}2$. This paradigm was motivated by irregular, often highly clumpy, morphologies and high velocity dispersions observed in the rest-frame UV with Hubble Space Telescope (HST) and ground-based facilities \citep[e.g.,][see \citealt{conselice14} and \citealt{forsterschreiber-wuyts20} for reviews]{forsterschreiber06,forsterschreiber09,forsterschreiber11,genzel06,genzel08,elmegreen07,conselice08,guo15,guo18,livermore15,wisnioski15,simons17}. However, with the advent of Atacama Large Millimeter/submillimeter Array (ALMA), it became feasible to observe the interstellar medium (ISM) of high-redshift galaxies in the [CII] fine-structure cooling line, which quickly revealed the abundance of dynamically cold disks all the way to $z \sim 6$ \citep[e.g.,][]{smit18,neeleman20,pensabene20,rizzo20,rizzo21,fraternali21,lelli21,tsukui-iguchi21,herreracamus22,posses23,romanoliveira23}. Nowadays, mounting evidence from JWST suggests that disk galaxies may in fact dominate out to $z \sim 8$, or universe age of $\lesssim 1$ Gyr \citep{ferreira22a,ferreira22b,jacobs22,naidu22,nelson22,robertson23}. Moreover, recent ``galactic archeology'' surveys of the chemistry and kinematics of nearby stars (facilitated by Gaia, APOGEE, LAMOST, and H3\footnote{\href{http://h3survey.rc.fas.harvard.edu/}{Hectochelle in the Halo at High Resolution}} among others) imply that the disk of our own Milky Way also formed quite early, within the first $1\text{--}2$ Gyr \citep{bk22,conroy22,rix22,xiang-rix22}. 

The formation of galactic disks has been a focus of much theoretical work over the past decades, both analytical \citep[e.g.,][]{fall-efstathiou80,ryden-gunn87,dalcanton97,mo-mao-white98} and numerical \citep[e.g.,][]{governato04,governato07,okamoto05,robertson06,scannapieco08,zavala08,brook12,bird13,bird21,fire,fire2,ubler14,agertz-kravtsov15,agertz-kravtsov16,genel15,christensen16,auriga,el-badry18,garrison-kimmel18,pillepich19,vintergatan1,vintergatan2,vintergatan3,gurvich22,hafen22,khoperskov22a,khoperskov22c,rodriguez-gomez22,vintergatan4,yu22,mccluskey23}. The classical analytical models (see the references above) usually consider disk formation as a result of the dissipational collapse of gas and assume that it retains the initial angular momentum acquired via torques during large-scale structure formation \citep[e.g.,][]{peebles69,doroshkevich70}. In this case, the angular momentum conservation leads to a linear correlation between galaxy and dark matter halo sizes. Observational studies indeed find such a linear correlation both for nearby galaxies \citep{kravtsov13} and as a function of redshift \citep{shibuya15}, with the normalization and scatter close to the expectations from the hierarchical structure formation. However, it was realized early on that the physics of galactic disk formation is more nuanced than the simple conservation of angular momentum during gas cooling and collapse.
For example, observational estimates of the retained angular momentum fractions are typically below unity and depend on the halo mass, galaxy baryonic fraction, and galaxy type \citep[e.g.,][]{dutton12,romanowsky-fall12,kauffmann15,mancera-pina21,romeo23}.

As another example, early galaxy formation simulations that modeled only gas cooling and gravity with weak or no star formation feedback struggled to produce realistic disks. Galaxies in such simulations suffered from angular momentum losses and became unrealistically compact, massive, and centrally concentrated---the problem dubbed ``angular momentum catastrophe'' in the early galaxy formation simulations literature \citep[e.g.,][]{navarro-white94,navarro95,navarro-steinmetz00}. This problem was partially alleviated in later simulations thanks to the improvements in resolution \citep[e.g.,][]{governato04,governato07} and hydrodynamical methods \citep[e.g.,][]{torrey12,vogelsberger12}, highlighting that the angular momentum loss is partially numerical in nature.

Another transformative change that helped to resolve the angular momentum catastrophe is the inclusion of strong galactic outflows driven by star formation and active galactic nuclei \citep[AGN; e.g.,][]{okamoto05,scannapieco08,zavala08,agertz-kravtsov15,agertz-kravtsov16,genel15}. Such outflows help to prevent or alleviate this problem by removing low-angular momentum gas from the galaxy and keeping the disk-to-spheroid ratio low, thereby suppressing disk instabilities that would otherwise transport angular momentum \citep[e.g.,][]{zavala08,brook12,agertz-kravtsov16,defelippis17}. In addition, reaccretion of ejected gas can bring high-angular momentum from the gaseous halo into the galaxy, thereby spinning up the disk \citep[e.g.,][]{sales12,ubler14,christensen16,defelippis17,fraternali17}.
Efficient feedback can also result in higher angular momentum of galaxies by suppressing star formation at the early stages dominated by the chaotic accretion of angular momentum and active galaxy mergers \citep[e.g.,][]{agertz-kravtsov16,garrison-kimmel18}.
Collectively, these results indicate that stellar and AGN feedback is crucial for the formation and survival of galactic disks, and the properties of simulated disks are sensitive to the details of feedback modeling \citep[e.g.,][]{okamoto05,zavala08,agertz-kravtsov15,agertz-kravtsov16,semenov21}.

More recently, \citet{stern19,stern20,stern21,stern23} put forward a model in which a dynamically cold thin disk tends to form after the inner circumgalactic medium (CGM) becomes thermally supported so that the cooling time is long compared to the infall time---the state that the authors dubbed ``inner CGM virialization'' (ICV). In such a state, the inflowing gas is accreted smoothly at subsonic velocities, with its angular momentum self-aligning and coherently joining the disk, while the outflows are confined to the disk by the pressure of the inner hot halo, leading to a steady evolution of the disk \citep{stern20,stern23,hafen22}. In contrast to such a ``hot-mode'' accretion, cold-mode supersonic inflow brings in clumpy material (as gas can cool quickly) in the form of cold streams with disordered angular momentum, perturbing or even disrupting the disk \citep[e.g.,][]{keres05,keres09,dekel-birnboim06,dekel09}. In addition, the clumpy inner halo provides low-density channels through which outflows can escape, resulting in significant variations of the ISM gas content and star formation rate (SFR), which also perturb the disk. A series of subsequent works indeed found a correlation between the onset of ICV and the settling of a thin disk in cosmological galaxy formation simulations \citep{stern21,hafen22,gurvich22,yu22}, although the causal connection has not yet been clearly demonstrated \citep[see][]{hopkins23disk}.

In another recent work, \citet{hopkins23disk} used a suite of galaxy simulations with variable parameters to show that the formation of a galactic disk is caused by a steepening of the gravitational potential in the region occupied by the cold gas. A shallow potential leads to efficient dispersal of the gas via various destructive modes, such as feedback-driven radial motions, while a steep centrally concentrated potential leads to suppression of such modes leaving behind only net rotation that results in disk formation. A centrally concentrated spherical contribution to the gravitational potential can also stabilize the disk via suppression of global bar instabilities \citep[e.g.,][]{efstathiou82,mo-mao-white98}. In this model, disk formation is a generic consequence of (mostly baryonic) material buildup in the halo center.
Interestingly, both this and the ICV picture described above have been investigated using the same simulation suite \citep[FIRE-2;][]{fire2,fire2-release}, yet the conclusions are qualitatively different.

Although substantial progress has been made in the past decade, it is still unclear what physical processes drive galactic disk formation. In this paper, we use the TNG50 cosmological simulation, part of the IllustrisTNG project, to gain further insights into this problem.
As was shown in previous works, TNG50 produces a population of galactic disks in reasonable agreement with observations, e.g., in terms of disk sizes as a function of galaxy mass and redshift \citep{pillepich19} and contains a substantial sample of MW and M31 analogs \citep{sotillo-ramos22,pillepich23}.
In the companion paper \citep[][hereafter \citetalias{semenov23a}]{semenov23a}, we also showed that TNG50 produces a subpopulation of MW-mass disks, which we dubbed ``early spin-up'' galaxies, that exhibit chemo-kinematical signatures of disk formation very close to the local spectroscopic and astrometric observations of nearby stars \citep{bk22}, implying analogous assembly history to the MW with early formation of the galactic disk and no subsequent destructive mergers. 

In this work, we investigate the physical processes operating in these chemo-kinematic MW analogs during the epoch when their disks are formed. In particular, we investigate the buildup of galactic angular momentum via the interplay between inflow and outflow and compare the TNG50 results with recent theoretical models of disk formation. 

The paper is organized as follows. In Section~\ref{sec:methods}, we briefly review the TNG50 simulation and outline our sample selection and key analysis details. In Section~\ref{sec:angmom}, we investigate the coevolution of the galactic angular momentum with that carried in and out by gas accretion and galactic outflows. Section~\ref{sec:potential} shows that the disk formation in TNG50 is correlated with a steepening of the central gravitational potential caused by the formation of a stellar bulge, in qualitative agreement with the \citet{hopkins23disk} results. In Section~\ref{sec:icv}, we show that disk formation is also correlated with a formation of a hot halo, in qualitative agreement with the \citet{stern19,stern20,stern21,stern23} model, with a distinction that hot- and cold-mode accretion coexist during disk formation. Section~\ref{sec:discussion} discusses our results, and Section~\ref{sec:summary} summarizes our conclusions.

%-----------------------------------------------------------------
%-----------------------------------------------------------------
\section{Methods}
\label{sec:methods}
%-----------------------------------------------------------------

%-----------------------------------------------------------------
\subsection{TNG50 Overview}
\label{sec:methods:tng50}
%-----------------------------------------------------------------

For our analysis, we use TNG50, the highest resolution run of the IllustrisTNG cosmological simulation suite \citep{springel18,pillepich18b,nelson18,marinacci18,naiman18,nelson19b}. Below is the brief summary of the key features of this simulation relevant to our work, in particular, the modeling of gas thermodynamics and feedback; for more detail, refer to \citet{pillepich19}, \citet{nelson19}, and the TNG method papers \citep{weinberger17,pillepich18}.

TNG50 is a simulation of a representative $\sim 50^3$ comoving Mpc$^3$ volume of the universe, assuming Planck XIII (\citeyear{planck-xiii}) cosmology carried out using the quasi-Lagrangian $N$-body and magnetohydrodynamic code Arepo \citep{arepo}. The mass resolution of dark matter particles is $4.5 \times 10^5\Msun$, while the mass of gas cells and star particles is kept within a factor of 2 from the target resolution, $8.5 \times 10^4\Msun$. The resulting median cell size in MW progenitor galaxies at $z=1$ is $\sim 80\text{--}100\pc$ \citep[see Figure~1 in][]{pillepich19}. The gravitational softening length for dark matter, stellar, and wind particles is 575 comoving pc at $z>1$ and 288 physical pc at $z<1$, while, for gas, the softening length varies with the cell size as $\epsilon_{\rm gas} = 2.5\;r_{\rm cell}$, where $r_{\rm cell}$ is the radius of a sphere with the same volume as the Voronoi cell.

TNG50 adopts the original Illustris model for gas cooling and heating \citep{vogelsberger13}. Radiative cooling accounts for primordial, Compton, and metal-line cooling, with an approximate model for self-shielding corrections in dense ISM. The gas is heated by the spatially uniform UV background, which is turned on at $z=6$ and thereafter follows the model of \citet{faucher-giguere09}, and by local AGN sources using the approximate method described in \citet{vogelsberger13}. 

Gas thermodynamics is followed at densities $n < 0.1 \cc$, whereas, at higher densities, it is replaced by the \citet{sh03} effective equation of state, with modifications described in \citet{vogelsberger13}. Such an approach is commonly used in large-scale cosmological simulations to handle the effects of local (i.e., unresolved) stellar feedback on the ISM pressure. One of the outputs of this subgrid model is the local SFR which is used to stochastically convert gas cells into star particles. The model parameters were calibrated against the observed Kennicutt--Schmidt relation from \citet{kennicutt98}, and, to speed up the calculations in TNG50, it was modified to more promptly convert gas denser than $24.4\cc$ to star particles \citep{nelson19}.

TNG50 includes galaxy-scale outflows driven by stellar and AGN feedback. Star formation-driven outflows are modeled via a kinetic wind approach: hydrodynamically decoupled wind particles are stochastically ejected from the ISM with a prescribed mass-loading at injection and redshift-dependent kick velocity proportional to the local velocity dispersion of dark matter particles with an imposed minimal value of $350 \kms$; wind particles are recoupled with gas when they reach the background density of $n = 2.5 \times 10^{-3} \cc$, typically within few kiloparsecs from the star-forming ISM \citep{pillepich18}. In contrast to the original Illustris model, the wind particles are launched locally isotropically, yet on the galaxy scale, a bipolar collimated wind develops preferentially perpendicular to the disk \citep{nelson19}.

Supermassive black holes producing AGN feedback are seeded when the host halo becomes sufficiently massive (when the mass of the friends-of-friends group reaches $5 \times 10^{10} \Msun$) and then they grow by accreting surrounding gas at the Edington-limited Bondi rate. Depending on the accretion rate ($\dot{M}_{\rm BH}$), the AGN feedback energy is injected in two modes: at high accretion rates, thermal energy is continuously injected in the surrounding gas at the rate of $0.02 \dot{M}_{\rm BH} c^2$, while, at low accretion rates, surrounding gas receives randomly oriented kicks with the typical kinetic energy injection rate of $\leq 0.2 \dot{M}_{\rm BH} c^2$ \citep{weinberger17}. In addition, the effect of radiative AGN feedback on the halo gas cooling and heating is treated using an approximate model described in Section~2.6.4 in \citet{vogelsberger13}.

%-----------------------------------------------------------------
\subsection{Milky Way-like Galaxy Sample and Analysis}
\label{sec:methods:sample-analysis}
%-----------------------------------------------------------------

\begin{figure*}
\centering
\includegraphics[width=\textwidth]{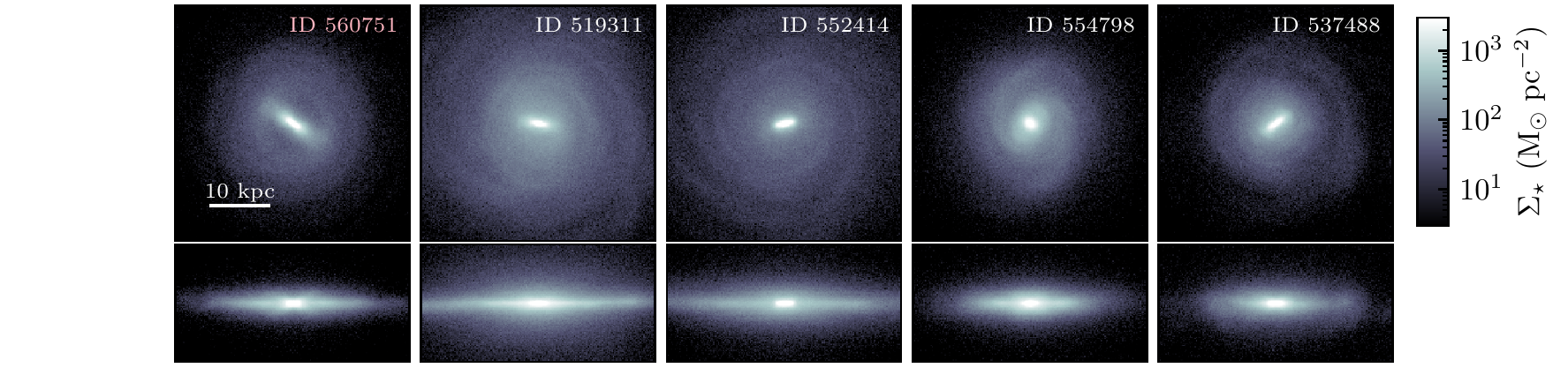}\\
\includegraphics[width=\textwidth]{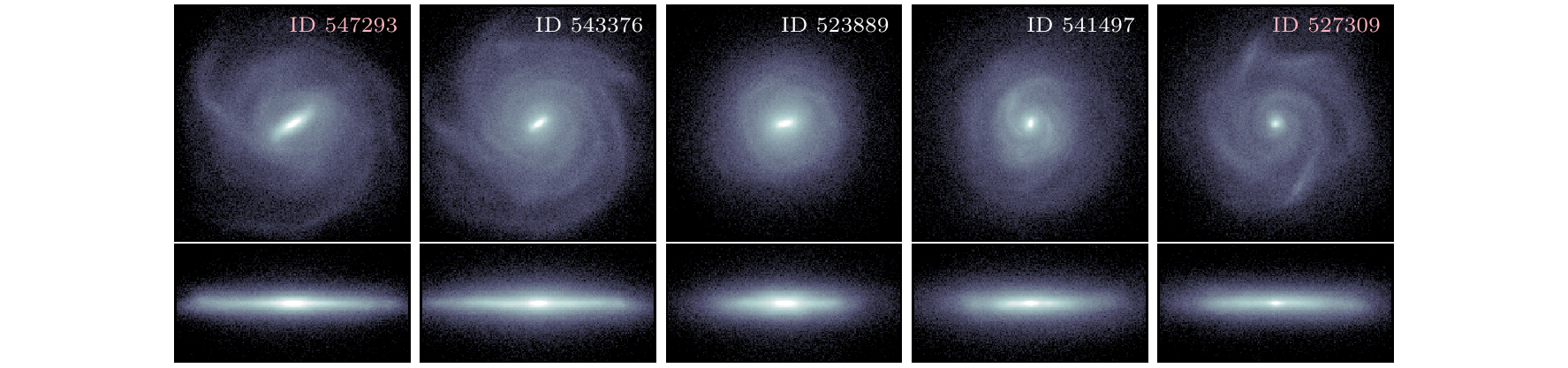}\\
\caption{\label{fig:sample} The sample of 10 early spin-up galaxies used for the analysis in this paper. These galaxies were selected from a larger pool of TNG50 MW-mass disk galaxies as the ones that form a stellar disk at the lowest metallicities, [Fe/H] $\lesssim -1$, in agreement with the local MW measurements \citep[see Figure~\ref{fig:vrot-stacked} below and][]{bk22,rix22}. \newtext{The maps are shown at redshift $z=0$; the color range in the bottom row is the same as in the top row as indicated by the color bar.} Host halo SubFind IDs are indicated in the corner of each panel, with red font indicating galaxies that we use as representative examples in Sections~\ref{sec:angmom}--\ref{sec:icv}: \underline{560751}, 519311, 552414, 554798, 537488, \underline{547293}, 543376, 523889, 541497, \underline{527309}.}
\end{figure*}

\begin{figure*}
\centering
\includegraphics[width=\textwidth]{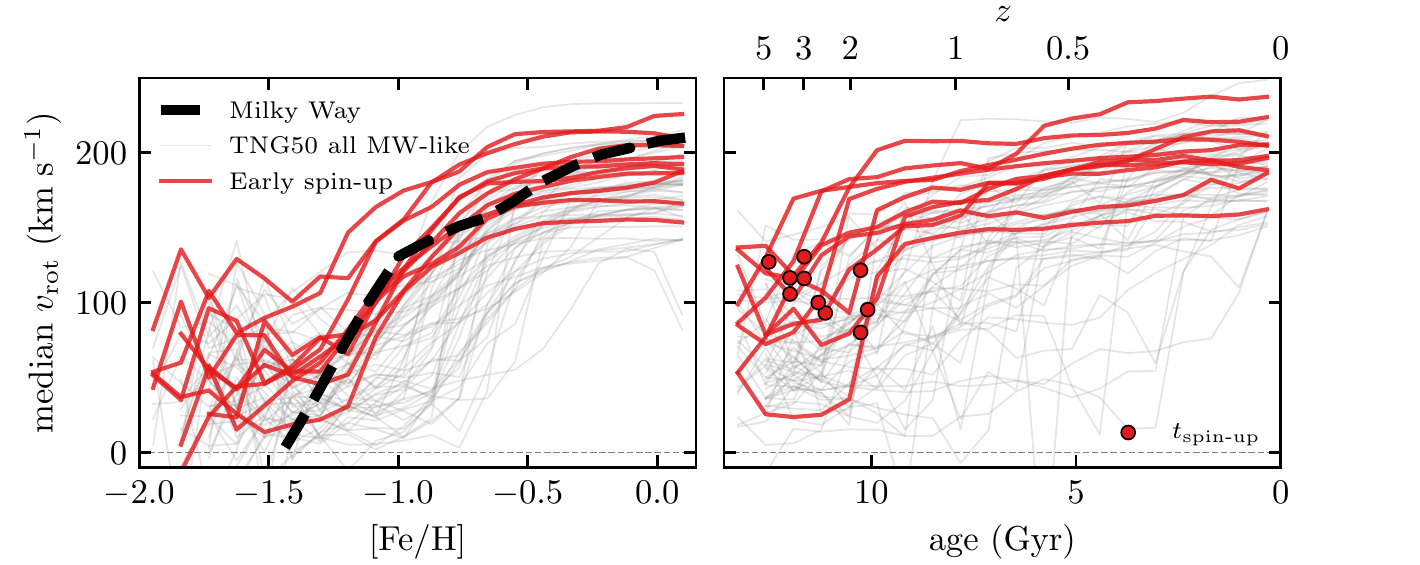}
\caption{\label{fig:FeH-vrot} Selection of galaxies based on the metallicity at which their stellar disks are formed. The panels show the running median of rotational velocity of star particles, $\vrot$, as a function of metallicity (that was renormalized to abundance-match the observed CDF as explained in Appendix A of \citetalias{semenov23a}) and age. The sharp rise of $\vrot$ imprints [Fe/H] and the age at the formation of the stellar disk \citep[][and \citetalias{semenov23a}]{bk22}. \newtext{The thick dashed black line shows the local observations from \citet{bk22}, while the thin dashed horizontal line at $\vrot = 0$ corresponds to no net rotation.} Red lines show the sample of TNG50 early spin-up galaxies that are most direct analogs of the MW in the $\vrot$ vs. [Fe/H] plane, which we focus on in this paper, while gray lines show the rest of MW-mass disk galaxies from TNG50 as described in the text. Circles in the right panel indicate the age at which median $\vrot$ reaches half of its final value, $\tspinup$, which we use as a proxy for the disk formation time in the rest of the paper.}
\end{figure*}

\begin{figure}
\centering
\includegraphics[width=\columnwidth]{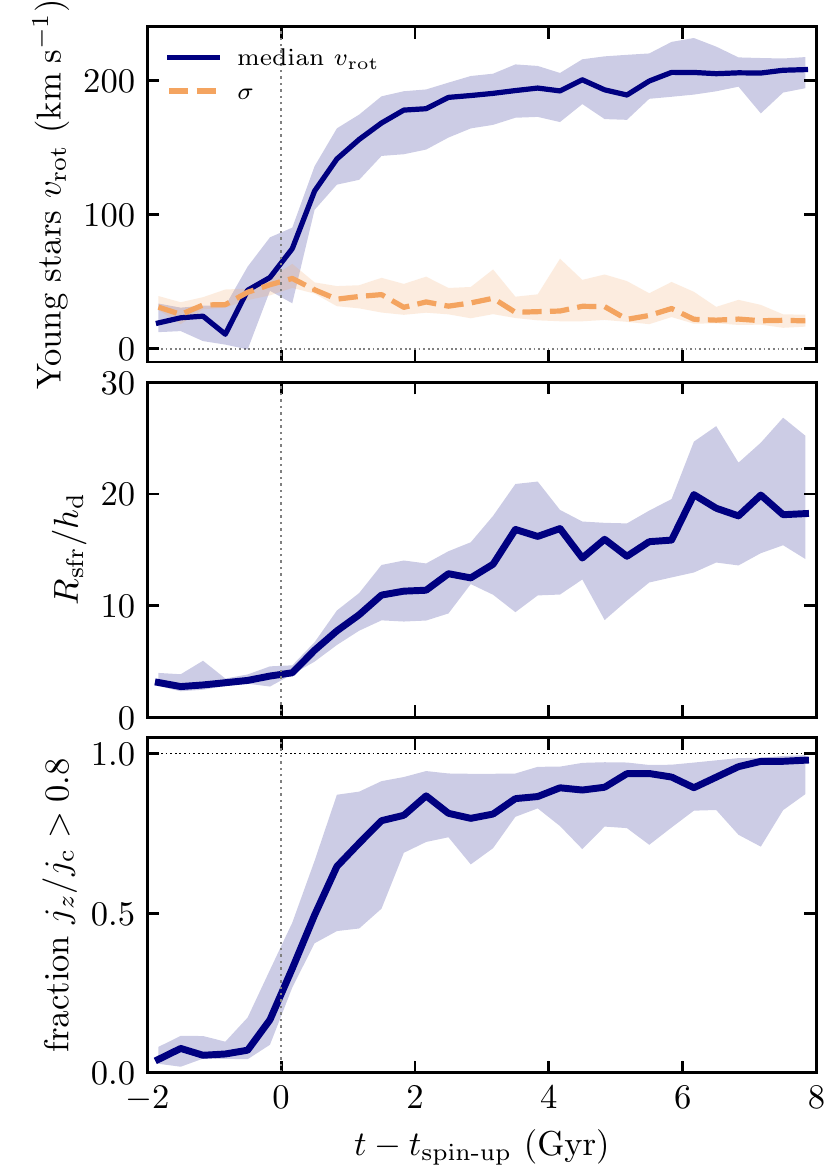}
\caption{\label{fig:vrot-stacked} \newtext{The evolution of various quantities describing the ``diskiness'' of the young, $<$100 Myr old, stellar population. {\bf Top:} median $\vrot$ and its dispersion (half of the 16--84 interpercentile range of $\vrot$). Unlike Figure~\ref{fig:FeH-vrot}, this plot shows $\vrot$ and $\sigma$ of stars at the moment when they were formed. {\bf Middle:} the ratio of the galaxy size ($\Rsfr$, twice the radius containing half SFR) to its thickness ($h_{\rm d}$ twice the mass-weighted average $z$ coordinate). {\bf Bottom:} the fraction of young stars on near-circular orbits with the orbital circularity of $j_z/j_{\rm c} > 0.8$.} The evolution tracks are stacked across our 10 early spin-up galaxies after offsetting them to the moment when the spin-up feature occurs in the age vs. median $\vrot$ of the final disk, $\tspinup$ (see the \newtext{right-hand} panel in Figure~\ref{fig:FeH-vrot}). \newtext{All these statistics confirm that a young stellar disk forms around $\tspinup$ making it a good proxy for the formation time of the disk.} }
\end{figure}

\emph{Galaxy sample.} For our analysis, we mainly focus on the sample of 10 ``early spin-up'' MW-like galaxies identified in \citetalias{semenov23a}. Figure~\ref{fig:sample} shows face-on and edge-on views of the stellar surface density in these galaxies at redshift $z=0$. This sample was chosen from a larger pool of 61 MW-mass disk galaxies that were selected based on their halo mass \newtext{($0.8 < M_{\rm 200c}/10^{12}\Msun < 1.4$)}, SFR ($\SFR > 0.2\Msunyr$), and rotational support of the young ($<100$ Myr) stellar disk at redshift $z=0$ ($\vrot/\sigma > 6$). 
The early spin-up galaxies were further selected as the ones that form a stellar disk at the lowest metallicities, [Fe/H] $\lesssim -1$, motivated by the observed relation between stellar [Fe/H] and rotational velocities \citep{bk22}. For this analysis, we recalibrated stellar metallicities predicted in TNG50 to exactly match the observed distribution as described in \citetalias{semenov23a}. \newtext{Our resulting sample contains 10 MW-analogs with halo virial masses---defined using the spherical overdensity from \citet{bryan-norman98}---ranging between $\Mvir \sim (0.9\text{--}15.6) \times 10^{12}\Msun$, virial radii $\Rvir \sim 250\text{--}300\kpc$, and virial velocities $\vvir = \sqrt{G\Mvir/\Rvir} \sim 125\text{--}150\kms$.}

This selection is demonstrated in Figure~\ref{fig:FeH-vrot}, which shows the median rotational velocity of stars near the solar circle (specifically, galactocentric $R = 5\text{--}11$ kpc and $|z| < 3$ kpc) as a function of stellar [Fe/H] and ages. The red lines show early spin-up galaxies, while the gray lines show the rest of the MW-mass disks in TNG50. The sharp increase of $\vrot$ at low metallicities and old ages imprints the formation of the galactic disk, or the ``spin-up.'' We identify the time of this spin-up, $\tspinup$, using the age at which median $\vrot$ reaches approximately half of its final value, and show it for the early spin-up galaxies in the right panel of the figure. \newtext{Note that, for each galaxy, $\tspinup$ is defined using only the distribution of stars in the final simulation output at redshift zero.} All 10 of these galaxies \newtext{show the spin-up feature at} $\gtrsim 10$ Gyr ago. 

\newtext{Figure~\ref{fig:vrot-stacked} demonstrates that $\tspinup$ reflects the moment in galaxy evolution when a stellar disk forms. The three panels show the instantaneous median $\vrot$ (rather than $z=0$ values as in the previous plot) and its scatter for the young ($<100$ Myr) star particles (top panel), the ratio of the radial extent of the same stellar population to its height (middle panel), and the fraction of young stars at near-circular orbits with the orbital circularity of $j_z/j_{\rm c} > 0.8$ as defined at the end of this section (bottom panel).} The evolutionary tracks are offset horizontally to align them at $\tspinup$. As the figure shows, a rotation dominated young stellar disk indeed forms around $\tspinup$, with the onset of disk formation at around $\tspinup-1$ Gyr and the end of the fast growth of $\vrot$ by approximately $\tspinup+2$ Gyr. At this point, the stellar disk of early spin-up galaxies acquires most of its final rotation velocity. Below, we will use $\tspinup$ as a proxy for the moment of disk formation. 

Here we focus on these early spin-up galaxies as they are the most direct analogs of the MW in terms of chemo-kinematic signatures, and they do not experience destructive mergers after disk formation enabling a cleaner investigation of this process and subsequent evolution. We have checked that our main conclusions remain qualitatively the same when the full sample of MW-mass disk galaxies is considered.
For more details about sample selection and the difference between early and late spin-up galaxies, see \citetalias{semenov23a}.

\emph{Galaxy size.} As a proxy for the size of the galaxy, we use two definitions: the size of a star-forming disk, $\Rsfr$, and the circularization radius of inflowing gas, $\Rcirc$. For each galaxy, $\Rsfr$ is defined as twice the radius containing half of the total SFR \newtext{averaged over 100 Myr}. The value of $\Rcirc$ depends on the net specific angular momentum of inflowing gas averaged over the entire halo, $|j_{\rm in}|$, and is defined as the radius of the circular orbit with the same specific angular momentum: $\Rcirc\;v_c(\Rcirc) = |j_{\rm in}|$, where $v_c(R)$ is the rotation curve of the galaxy. We use $\Rcirc$ for the comparison with the \citet{stern19} model below; we find that these two definitions agree on average (see Figure~\ref{fig:vc-stack}). \newtext{By visually inspecting galaxies from our sample, we find that $\Rsfr$ and $\Rcirc$ trace the extent of the disk better than, e.g., stellar half-mass radii, which can be biased low in the presence of strong bulges.}

\emph{Galaxy orientation.} Rotational velocity\newtext{, $\vrot$,} and the angular momentum vector\newtext{, $(j_x,j_y,j_z)$,} are defined with respect to the frame in which the $z$-axis is directed along the angular momentum vector of the existing stellar population. \newtext{$\vrot$ for each particle is defined as the tangential component of the particle velocity in that frame. In our analysis, we use the median value of $\vrot$ and its scatter (half of the 16--84 interpercentile range) as measures of net rotation and velocity dispersion, respectively. Further,} the ratio $j_z/|j|$ for any cell or particle or a collection of such will reflect the level of angular momentum alignment with the existing stellar disk, which we use in our analysis to quantify the coherence of the angular momentum brought in and out by accretion and galactic winds. 

\emph{Orbital circularity.} In Figure~\ref{fig:jzjc-spinup} below, we use the distribution of orbital circularities of star particles, $j_z/j_{\rm c}$, to kinematically separate the stellar population into bulge and disk. $j_{\rm c}$ of a particle is the specific angular momentum of the circular orbit having the same energy as the particle energy, $E = \phi + |v|^2/2$, with $\phi$ and $|v|$ being the gravitational potential and the magnitude of particle velocity with respect to the galaxy, respectively. To calculate $j_{\rm c}$, we use a spherically averaged $\phi$ and invert the relation between orbital energy and radius, $E_{\rm c}(R_{\rm c}) = \phi(R_{\rm c}) + v_{\rm c}(R_{\rm c})^2/2$, with $v_{\rm c}^2 = R\;d\phi/dR$, to compute $j_{\rm c} \equiv R_{\rm c} \; v_{\rm c}(R_{\rm c})$ at $R_{\rm c}(E_{\rm c} = E)$. By construction, $j_z/j_{\rm c}$ can take values from 1 (corotation with the galaxy on a circular orbit) to $-1$ (counterrotation), with 0 corresponding to the motion along radial or polar orbits.

%-----------------------------------------------------------------
%-----------------------------------------------------------------
\section{Buildup of Galactic Angular Momentum}
\label{sec:angmom}
%-----------------------------------------------------------------

\begin{figure}
\centering
\includegraphics[width=\columnwidth]{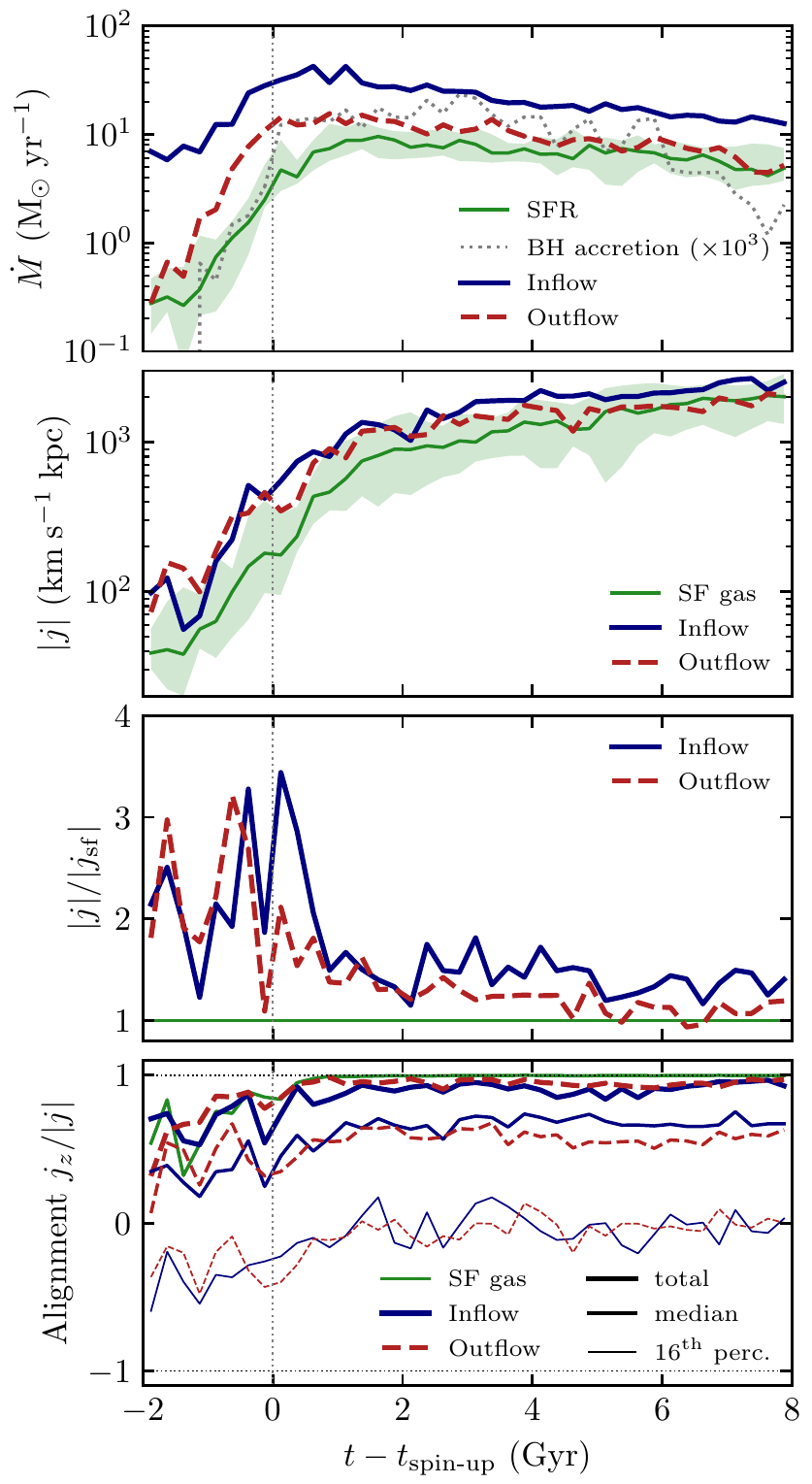}
\caption{\label{fig:jz-stacked} The evolution of angular momentum contained in inflowing, star-forming, and outflowing gas. \newtext{The evolution tracks are aligned at $\tspinup$ (see Figure~\ref{fig:vrot-stacked} and related text).} The panels show from top to bottom: (1) inflow (gas with radial velocity $v_r < -0.5\,\vvir$) and outflow ($v_r > 0.5\,\vvir$) rates at $0.2\,R_{\rm vir}$, together with the SFR and black hole accretion rate; (2) specific angular momentum of inflowing, star-forming, and outflowing gas; (3) excess of specific angular momentum of inflowing and outflowing gas relative to that of the star-forming gas; (4) alignment of angular momentum in different components with that of the stellar disk. \newtext{The quantities shown in the plot all exhibit a factor of $\sim 2$ galaxy-to-galaxy variation; for reference, shaded regions indicate 16--84 interpercentile ranges of SFR and $|j|$ of star-forming gas.} The growth of the angular momentum of the star-forming disk is facilitated by the inflow of high-$|j|$ material; it is rapid early on when the excess is large and saturates at later times when $|j|$ removed by outflows and brought back by inflows become comparable to that of the galaxy.
\newtext{In this plot, we consider the inflow and outflow through a spherical shell at $0.2\,R_{\rm vir}$ but we also find analogous coevolution throughout the halo including closer to the galactic disk at distances $\Rsfr$ and $\Rcirc$.}
}
\end{figure}

The angular momentum of star-forming gas is arguably the key property controlling the formation and evolution of galactic disks (see Introduction). 
In this section, we investigate how the angular momentum of the forming star-forming disks is built up as a result of the interplay between inflows and outflows bringing angular momentum into and out of the galaxy (Section~\ref{sec:angmom:winds}). We also show that the distributions of angular momentum alignment in the inflowing and outflowing gas are remarkably similar, indicating either efficient exchange of angular momentum, wind recycling, and/or a similarity in the evolution ways of both components (Section~\ref{sec:angmom:alignment}). Finally, we also investigate the variation in the orientation of the angular momenta of the disk, inflow, and outflow (Section~\ref{sec:angmom:frame}). Overall, the results of this section suggest that the outflows can play an important role in mediating the angular momentum exchange between the inflow from the intergalactic medium (IGM) and the galaxy. We further discuss these results in Section~\ref{sec:discussion:winds}.

%-----------------------------------------------------------------
\subsection{The Interplay between Inflow and Outflow}
\label{sec:angmom:winds}
%-----------------------------------------------------------------

Figure~\ref{fig:jz-stacked} shows the buildup of angular momentum in our sample of early spin-up galaxies around the moment of disk formation ($\tspinup$; see Section~\ref{sec:methods:sample-analysis}). The top panel shows the evolution of gas accretion and expulsion rates through a spherical shell at $0.2\,R_{\rm vir}$ \newtext{with the width of $\sim 0.05\,R_{\rm vir}$} together with the SFR and central black hole accretion rate (proxies for stellar and AGN feedback activity). 
All curves have qualitatively similar shapes: rapid rise before and during disk formation (vertical dotted line) and saturation with a slow decline at later times ($>1\Gyr$ after disk formation). This similarity is not surprising because these rates are causally related: the inflow rate increases the gas density and the total gas mass available for star formation and black hole feeding, which in turn drives outflows.
The overall shape also reflects the two modes of galaxy mass assembly: early rapid buildup followed by a more gradual evolution (see, e.g., Figure~7 in \citetalias{semenov23a}).

The second panel from the top shows the evolution of the specific angular momentum contained in the inflowing (blue), star-forming (green), and outflowing gas (red). Again, all three components show qualitatively similar tracks: the angular momentum quickly increases over the period from $\sim 2\Gyr$ before to $\sim 2\Gyr$ after the disk is formed and then gradually increases at a slower rate (i.e., with a longer $e$-folding time).
It is particularly interesting that the inflows and outflows carry approximately the same specific angular momentum. 
\newtext{Although here we consider the inflow and outflow through a spherical shell at $0.2\,R_{\rm vir}$, we also find analogous coevolution throughout the halo including closer to the galactic disk at distances $\Rsfr$ and $\Rcirc$.}

Apart from being qualitatively similar, evolutionary tracks of $|j|$ also show quantitative differences that provide insights into how galactic disks form and grow. The third panel from the top shows the angular momenta of inflowing and outflowing material normalized by that of the star-forming gas. 
Prior to disk formation (to the left of the vertical dotted line), the specific angular momentum contained in the inflow and outflows is $\sim 2\text{--}3$ times higher than that of the galaxy.
The \emph{mass} inflow rate, however, is several times higher than the outflow rate (see the top panel), leading to the net inflow of high-angular-momentum gas, which causes the rapid increase of the galactic $|j|$ and therefore the spin-up of the disk at $\tspinup$ (recall Figure~\ref{fig:vrot-stacked}). 
At later times, $\gtrsim 2\Gyr$ after disk formation, this process saturates, as the specific angular momentum removed by outflows and brought back by accretion becomes comparable to that contained in the star-forming disk. As a result, the net specific angular momentum (green line in the second panel) and rotational velocity of stars (blue line in Figure~\ref{fig:vrot-stacked}) settle down and continue increasing only moderately.

%-----------------------------------------------------------------
\subsection{Angular Momentum Alignment}
\label{sec:angmom:alignment}
%-----------------------------------------------------------------

\begin{figure*}
\centering
\vspace{1em}
{\large \bf Alignment of total angular momentum, $\langle j_z \rangle/|j|$}\\
\includegraphics[width=0.3559\textwidth]{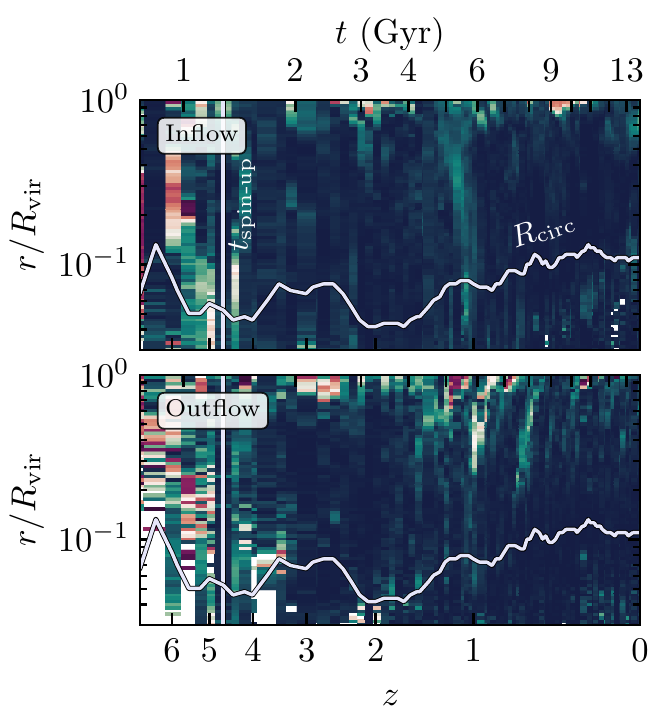}%
\includegraphics[width=0.2880\textwidth]{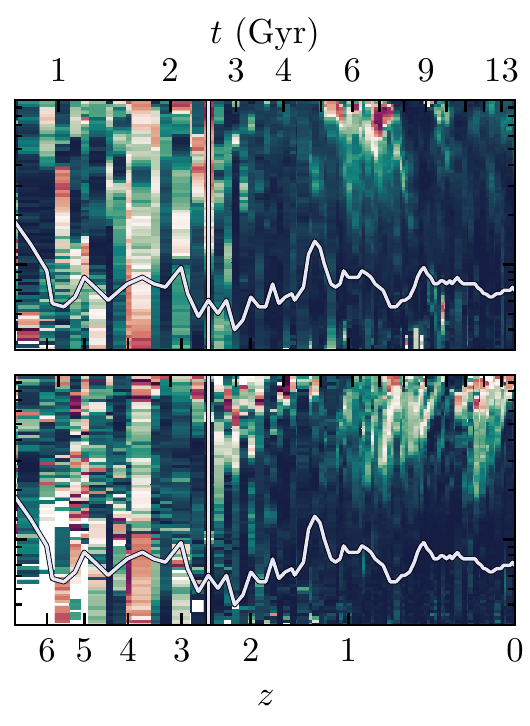}%
\includegraphics[width=0.3559\textwidth]{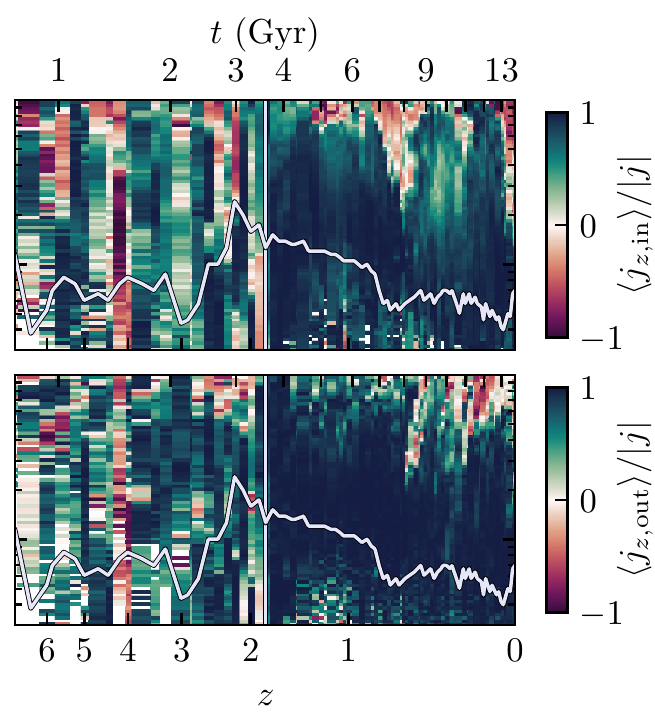}\\
\vspace{1em}
{\large \bf Misalignment, 16$^\text{th}$ percentile of $j_z/|j|$}\\
\includegraphics[width=0.3559\textwidth]{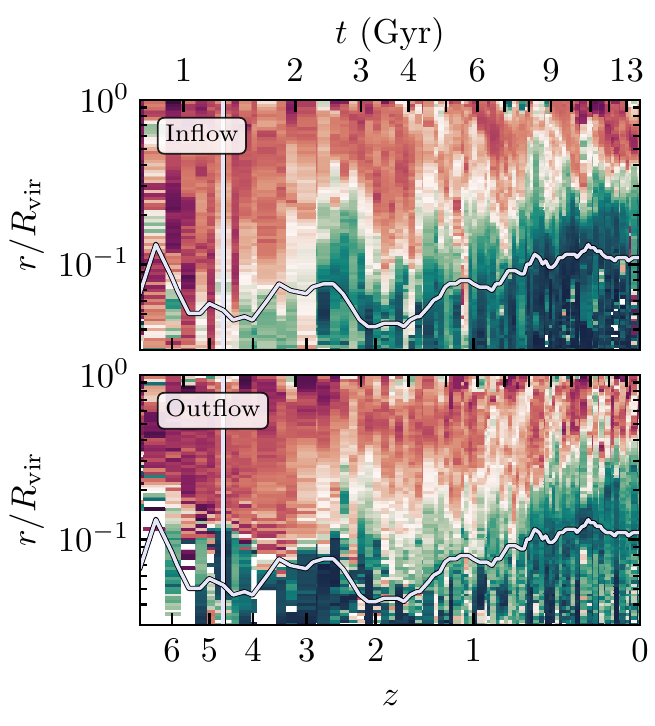}%
\includegraphics[width=0.2880\textwidth]{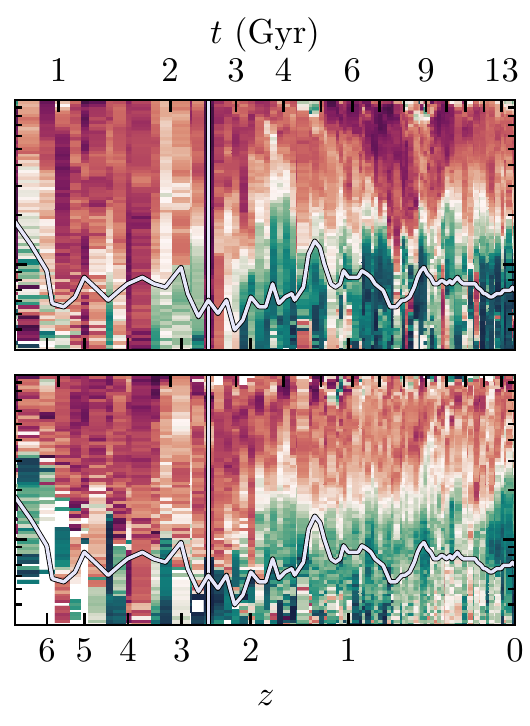}%
\includegraphics[width=0.3559\textwidth]{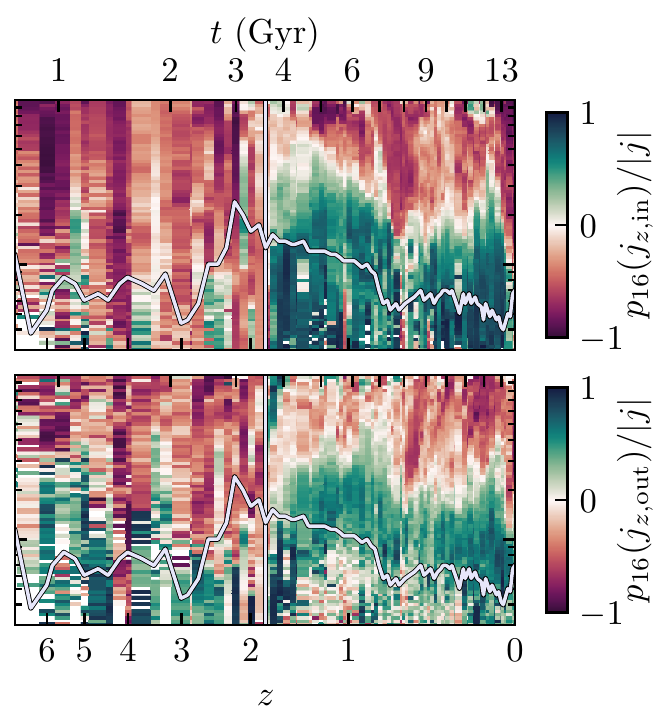}\\
\caption{\label{fig:jz} Examples of evolution of angular momentum alignment profiles in three early spin-up galaxies. The maps show the degree of gas alignment with the stellar disk, with green and red indicating co- and counterrotation, respectively. The top set of panels shows the alignment of the \emph{net} angular momentum at a given radius, while the bottom set of panels shows the 16$^{\rm th}$ percentile of alignment for individual gas cells; i.e., how misaligned the gas can be at a given radius. The top and bottom rows in each panel show inflowing and outflowing gas, respectively. Vertical lines show the moment at which the spin-up feature occurs in stellar ages (see the \newtext{right-hand} panel in Figure~\ref{fig:FeH-vrot}), while the horizontal curve shows the evolution of the circularization radius as a proxy for the galaxy size ($R_{\rm circ}$; see Section~\ref{sec:methods:sample-analysis}). Both inflowing and outflowing gas are highly aligned with the stellar disk after the disk is formed, and the distribution of alignment is similar between the two components.}
\end{figure*}

\begin{figure*}
\centering
\includegraphics[width=0.3559\textwidth]{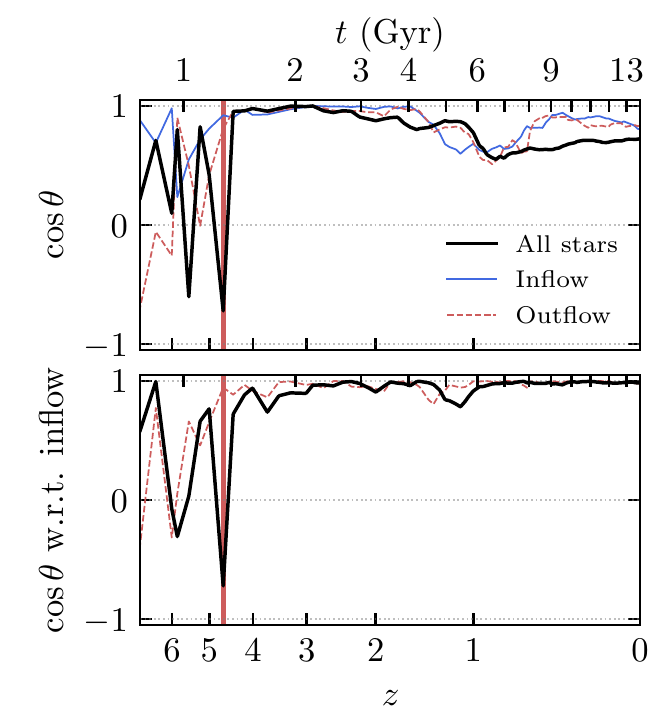}%
\includegraphics[width=0.2880\textwidth]{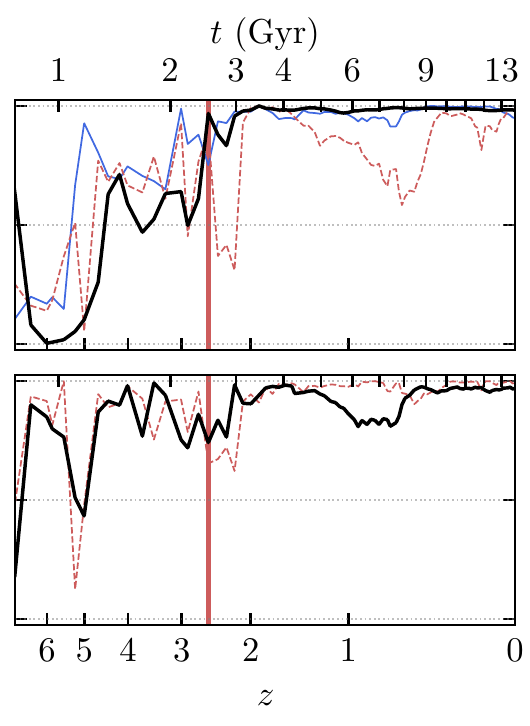}%
\includegraphics[width=0.3559\textwidth]{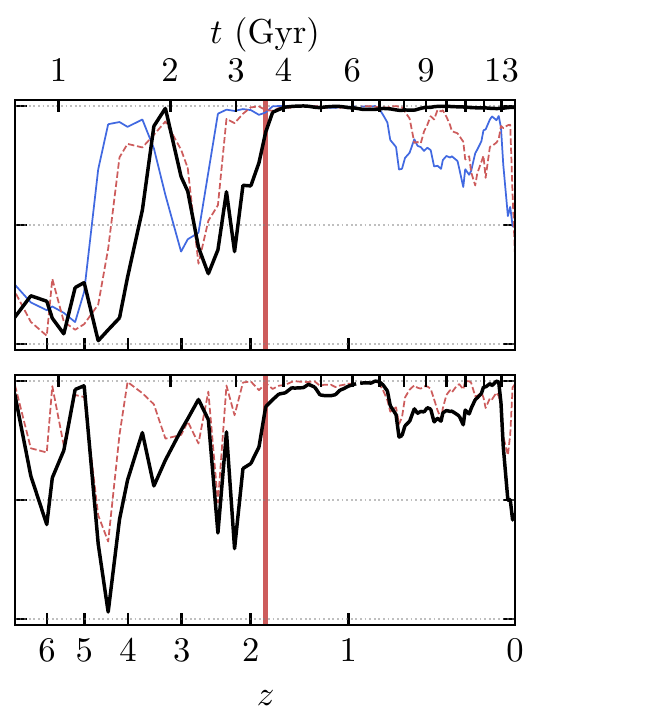}\\
\caption{\label{fig:alignment-examples} Evolution of the net angular momentum orientation of the stellar disk (thick black line), inflow (thin blue line), and outflow (dashed red line) for the three example halos from Figure~\ref{fig:jz}. {\bf Top:} absolute orientation; i.e., the cosine of the angle between the instantaneous angular momentum with its direction shortly after disk formation, at $\tspinup + 1$ Gyr. {\bf Bottom:} orientation relative to the inflow; i.e., the cosine of the angle between the net angular momenta of the inflow and, respectively, stellar disk and outflows. The values of $\cos \theta = 1$, $0$, and $-1$, correspond to perfectly aligned, perpendicular, and oppositely directed angular momenta, respectively. Vertical red lines mark the moment of disk formation, $\tspinup$. Before disk formation, the angular momenta of all three components exhibit significant variation in their directions, while, after the disk forms, these variations are reduced. This result indicates that the sharp transition from misaligned inflow and outflow before disk formation to strong alignment at later times in Figure~\ref{fig:jz} is caused by both the change in the reference orientation of the galaxy and the inflow and outflow themselves before disk formation.}
\end{figure*}

In addition to the net angular momentum contained in the inflowing and outflowing material, its alignment with that of the galactic disk is crucial. Although a transient gas disk might form as a result of spontaneous coplanar accretion, to grow a stellar disk, high-angular-momentum material must be fed continuously with a coherent orientation \citep[e.g.,][]{sales12,meng-gnedin21,kretschmer22}. 
The bottom panel in Figure~\ref{fig:jz-stacked} demonstrates that, indeed, the angular momentum of both inflowing and outflowing gas is highly aligned with the galactic disk. The green line shows the orientation of the net angular momentum of star-forming gas with respect to the existing stellar disk (see Section~\ref{sec:methods:sample-analysis} for details), while thick lines show the alignment of the total angular momentum contained in the inflowing and outflowing material. Before a stellar disk has formed, all three components have angular momentum that is somewhat aligned with the existing stellar population ($j_z/|j| \sim 0\text{--}0.8$). As we show below, this is due to both the stochasticity in the inflow directions and the lack of a well-defined reference orientation at early times. After the disk has formed, however, all three components are highly aligned with the stellar disk, $j_z/|j| > 0.8$. 

In addition to having similar alignment of the total angular momentum, the inflowing and outflowing gas have remarkably similar distributions of the angular momentum alignment of individual gas cells. Both components have similar median and 16$^{\rm th}$ percentile of $j_z/|j|$ (lines of different thickness in the bottom panel of Figure~\ref{fig:jz-stacked}). The median $j_z/|j|$ of both inflow and outflow grows from $\sim 0$ (i.e., random orientation) to $\sim 0.6$ during the formation of the disk and remains constant thereafter. This value is somewhat smaller than the alignment of the total momentum, implying that some of the misaligned angular momenta of individual particles cancels out when they add up in the total angular momentum. The 16$^{\rm th}$ percentile  saturates at the level of $j_z/|j| \sim 0$ indicating that the majority of gas (84\% of cells) in the inflow and outflow corotates with the disk; i.e., has $j_z/|j| > 0$. 

The above results show that the inflow and outflow corotate with the disk and have similar distributions of angular momentum directions at $0.2\, \Rvir$. Figure~\ref{fig:jz} further demonstrates that inflow and outflow exhibit similar distributions of $j_z/|j|$ not only at $0.2\, \Rvir$ but throughout the entire halo. The top set of panels shows the evolution of the orientation of the net angular momentum, while the bottom set shows the 16$^{\rm th}$ percentile of $j_z/|j|$ for three representative halos from our sample. The halos are ordered according to the time of galactic disk formation, $\tspinup$, which is indicated with the vertical line. For reference, the curve at the bottom of each panel shows the circularization radius as a proxy for the size of the galaxy (see Section~\ref{sec:methods:sample-analysis} for the definition).

A comparison of the top (inflow) and bottom (outflow) rows in each set of panels shows that the total alignment and 16$^{\rm th}$ percentiles of $j_z/|j|$ are quite similar for both components throughout the halo ($y$-axis) and at all times ($x$-axis). The total alignment of both components is remarkably high $\gtrsim 0.8$ especially after the disk is formed (vertical lines). The prominent red streaks at $r>0.3\;\Rvir$ reflect misaligned mergers and passerby galaxies. 

The 16$^{\rm th}$ percentile of $j_z/|j|$ shown in the bottom set of panels can be thought of as a measure of the width of the $j_z/|j|$ distribution, or equivalently, how misaligned inflowing and outflowing gas can be at a given distance from the galaxy. Three conclusions follow from this set of panels. (i) Again, the maps are remarkably similar between the inflows and outflows for a given halo, indicating that the distributions of the angular momentum in the inflowing and outflowing gas are quite similar. (ii) The inflowing and outflowing angular momentum close to the disk (horizontal curves at the bottom of each panel) are misaligned before disk formation and become aligned only after the disk is formed, implying that the strong alignment of inflowing angular momentum might not be a prerequisite but rather a consequence of galactic disk formation. (iii) The alignment of inflowing angular momentum with the disk (transition from red to green color) occurs at a significant distance from the disk, $\sim 3\text{--}5$ times the circularization radius.

A possible interpretation of the above results, in particular, the similarity between the angular momenta of inflow, outflow, and star-forming gas, is that galactic outflows may mediate angular momentum exchange between the inflow and the disk. Being driven by star formation and AGNs, outflows originate from the galactic disk, and therefore, they can bring the angular momentum of the disk into the halo and exchange it with the inflowing material and bring it back via outflow recycling. This implies that \emph{the exchange of angular momentum between the inflow and the disk can be facilitated by the outflows and occur within the halo, or corotating CGM, before the inflowing material reaches the disk}. We discuss this scenario further in Section~\ref{sec:discussion:winds}.

One remarkable feature of the distributions shown in Figure~\ref{fig:jz} is the sharpness of the transition from misaligned inflow and outflow before the disk formation to strong alignment after the disk is in place. In the next subsection, we show that this transition is a result of both the settling of disk orientation and the more coherent alignment of the gas inflow and outflow after this settling.

%-----------------------------------------------------------------
\subsection{Variation in Disk, Inflow, and Outflow Orientations}
\label{sec:angmom:frame}
%-----------------------------------------------------------------

The alignment of the inflow and outflow in the previous subsection is defined with respect to the instantaneous orientation of the stellar disk (see Section~\ref{sec:methods:sample-analysis} for the definition). Therefore, the temporal variations of the alignment can be due to either variation in the gas flow direction, or changes in the stellar disk orientation, or a combination thereof.

To disentangle these factors, Figure~\ref{fig:alignment-examples} shows the variation of the orientation of the net angular momentum of the stellar disk, inflows, and outflows with respect to their orientations when the disk is established (1 Gyr after $\tspinup$; top panels), and the relative orientation of the disk with respect to the direction of the net angular momentum of the inflowing gas (bottom panels).

As the top panels show, prior to disk formation, all three components randomly change their orientations. The change in orientation of the stellar disk lags behind that of the infalling gas as newly accreted gas defines the orientation of the young stellar disk, which, if it remains coherent, eventually becomes the new orientation of the stellar disk. The angular momentum of outflows follows that of the young stellar disk and, therefore, is offset between the inflow and stellar disk (this is especially clear in the top right panel before the disk formation).

After disk formation (to the right from the vertical red lines), the orientation of the stellar disk settles and changes only mildly. The orientation of the inflow and outflow also exhibit smaller variations after disk settling, although, in the middle and right panels, the variations are significantly stronger than that of the stellar disk. These variations are caused by merger and flyby events seen as the prominent red streaks in the top set of panels of Figure~\ref{fig:jz}; the stellar disk in these cases survives these events \citep[see also][]{dillamore22,sotillo-ramos22}.

The bottom panels in Figure~\ref{fig:alignment-examples} show the instantaneous orientation of the stellar disk and outflows with respect to the inflow. This relative orientation shows trends similar to those in the top panel: the variations are stronger prior to disk formation and become smaller after the stellar disk settles.

Altogether, the results shown in Figure~\ref{fig:alignment-examples} imply that the transition from the incoherent early accretion and outflow to significantly stronger alignment after disk formation results from both the variation of the reference orientation of the galaxy and stochastic changes of the inflow and outflow directions at early times. Indeed, not only the stellar disk but also inflows and outflows show significant variations of orientation at early times (top panels in the figure). The relative orientations also vary significantly at early times (bottom panels in the figure). After the disk settles, its orientation changes only slightly, and both inflow and outflow are on average aligned with that orientation.

%-----------------------------------------------------------------
%-----------------------------------------------------------------
\section{Coevolution of Disk and Gravitational Potential}
\label{sec:potential}
%-----------------------------------------------------------------

\citet{hopkins23disk} investigated galactic disk formation in a suite of galaxy formation simulations with varying shapes of gravitational potential and multiple other parameters. The authors conclude that the susceptibility of gas to disk formation mainly depends on the steepness of the gravitational potential within the region occupied by the gas. A steep and centrally concentrated potential plays two roles: (i) it defines a center around which a galactic disk can form, and (ii) it efficiently suppresses radial motions of gas leaving only the tangential velocity component; i.e., the net rotation. A centrally concentrated spherical contribution to the gravitational potential can also stabilize the disk via suppression of global bar instabilities \citep[e.g.,][]{efstathiou82,mo-mao-white98}.

In this section, we show that the disk formation in TNG50 is consistent with the above scenario. The formation of the disk coincides with the steepening of the central potential on a few Gyr-long timescale (Section~\ref{sec:potential:steepening}). This steepening is caused by the formation of the stellar bulge from the remnant of the pre-spin-up protogalaxy (Section~\ref{sec:potential:bulge}). Note, however, that such a correlation may not necessarily imply causation as significant steepening occurs not before but after the onset of disk formation $\sim 1$ Gyr prior to $\tspinup$. At the same time, such a steepening can play an important role during disk formation even if the formation of the disk itself is not caused by the change in the potential shape (see Section~\ref{sec:discussion:disk-formation} for further discussion).

%-----------------------------------------------------------------
\subsection{Correlation of Potential Steepening and Disk Spin-up}
\label{sec:potential:steepening}
%-----------------------------------------------------------------

\begin{figure}
\centering
\includegraphics[width=\columnwidth]{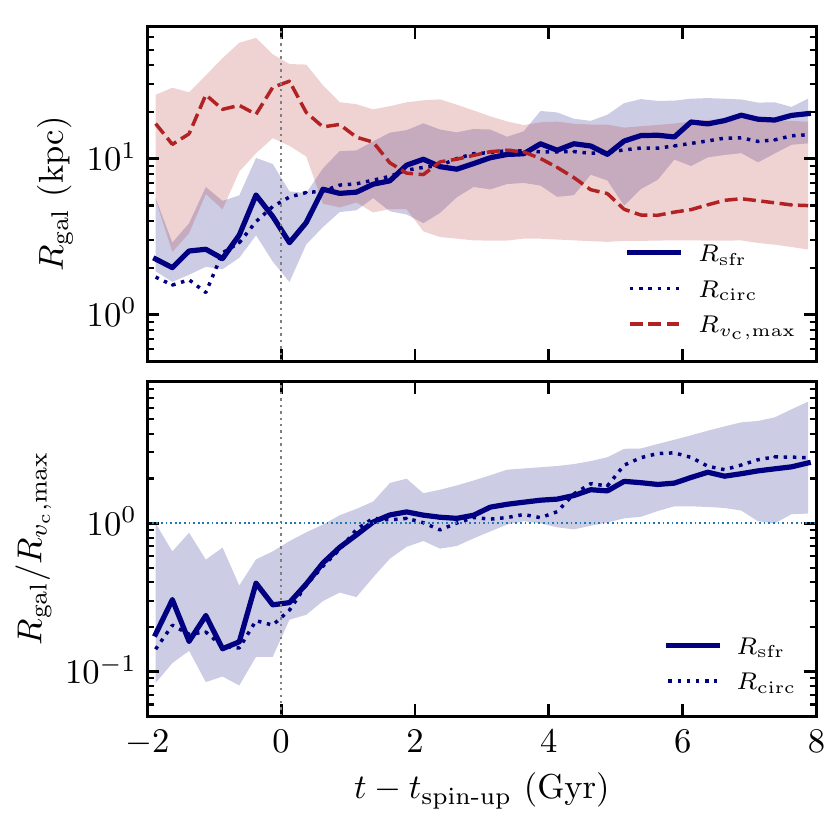}
\caption{\label{fig:vc-stack} Evolution of the steepness of the gravitational potential in the region occupied by the galaxy for the sample of our early spin-up galaxies. {\bf Top:} comparison of the galaxy size (blue lines) with the radius at which rotational velocity reaches its maximum, $\Rvmax$. For the galaxy size, we use two definitions: the size of the star-forming disk ($\Rsfr$, twice the radius containing half of the total SFR), and the circularization radius ($\Rcirc$, see Section~\ref{sec:methods} for the definition)---both agree quite well. {\bf Bottom:} The ratio of the galaxy size to $\Rvmax$. This ratio increases monotonically with time implying that the gravitational potential becomes more centrally concentrated within the region occupied by the galaxy.}
\end{figure}

\newtext{We follow \citet{hopkins23disk} and characterize the steepness of the potential using the disk size and the radius at which the rotation curve reaches its peak, as is done by these authors.
Figure~\ref{fig:vc-stack} shows the evolution of these characteristic spatial scales in our sample of early spin-up galaxies.} The top panel compares the galaxy size with the radius at which the rotational curve reaches its maximum, $\Rvmax$. As a proxy for the galaxy size, we use the size of the star-forming disk ($\Rsfr$, defined as twice the radius containing half of the total SFR) and the circularization radius of inflowing gas ($\Rcirc$, defined as the radius of a circular orbit corresponding to the specific angular momentum contained in the inflowing halo gas; see Section~\ref{sec:methods:sample-analysis}). These two definitions of the galaxy size closely agree after a disk has formed, while, before disk formation, $\Rsfr$ is a factor of 2--3 larger.

As the figure shows, the size of the galaxy monotonically increases, while $\Rvmax$ increases at early times, reaches a peak around the epoch of disk formation, and decreases afterward. The ratio of these two scales reflects the steepness of the gravitational potential within the galaxy, and, according to \citet{hopkins23disk}, a disk should form when this value reaches a certain value, meaning that the potential becomes sufficiently steep. 
As the second panel shows, the ratio of galaxy size to $\Rvmax$ monotonically increases, implying that the gravitational potential becomes progressively more concentrated. Comparison with Figure~\ref{fig:vrot-stacked} suggests that disk formation sets in when $\Rsfr/\Rvmax \approx 0.2\text{--}0.3$ and continues until $\Rsfr/\Rvmax \approx 1$ reached $2 \Gyr$ after $\tspinup$. 

\begin{figure*}[t]
\centering
\includegraphics[width=0.49\textwidth]{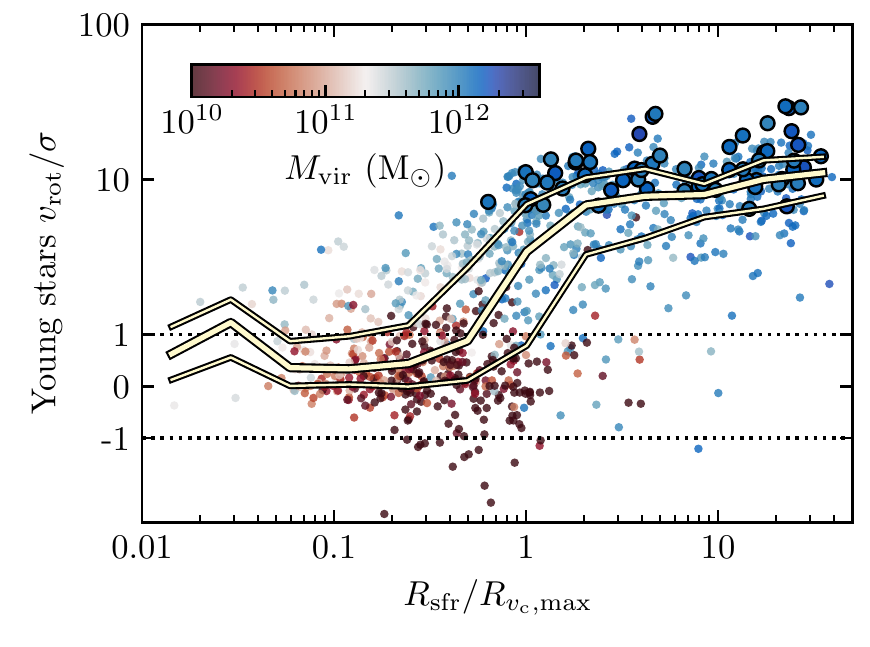}%
\includegraphics[width=0.49\textwidth]{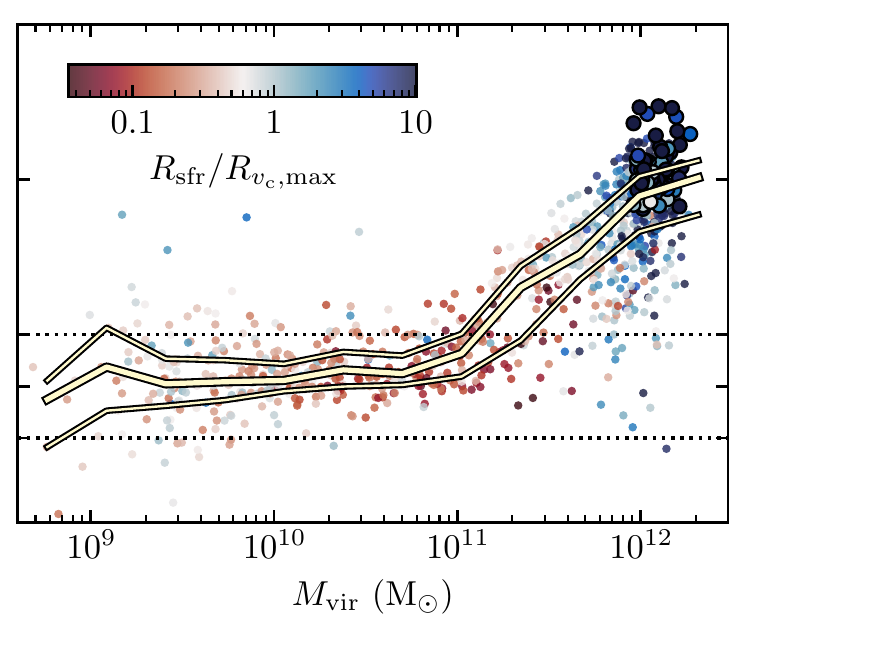}
\caption{\label{fig:rfrac-vc} Relation between the rotational support of the young ($<100$ Myr old) stellar disk, $\vrot/\sigma$, the ratio $\Rsfr/\Rvmax$ that describes the steepness of the gravitational potential in the region occupied by the star-forming disk, and the virial mass of the host halo, $\Mvir$. The left and right panels show different projections of this relation: $\vrot/\sigma$ vs. $\Rsfr/\Rvmax$ and $\Mvir$, respectively, colored by the third property. The points show all 61 of our MW-mass galaxies identified in TNG50 across available snapshots spanning the range of redshifts from $z \approx 12$ to $0$. The final locations of the galaxies are marked with \newtext{large circles}. \newtext{Yellow} lines show the running medians and 16--84 interpercentile ranges. The scale of the $y$-axis is linear for $|\vrot|/\sigma<1$ (i.e., between the dotted lines) and logarithmic outside of this range. Although the scatter is large, the rotational support of young stellar disks rapidly increases as the central potential becomes sufficiently steep ($\Rsfr/\Rvmax \gtrsim 0.5$), in qualitative agreement with the picture of \citet{hopkins23disk}. Interestingly, disks tend to form when the host halo becomes more massive than $\sim (1\text{--}2) \times 10^{11} \Msun$; it is not clear, however, whether this threshold is caused by the central potential steepening or other factors, e.g., a hot halo formation (see Section~\ref{sec:discussion:disk-formation}). The values of the median $\vrot$ for young stars are negative for some of the galaxies because the frame is aligned with the angular momentum of all stars in the galaxy.}
\end{figure*}

The correlation between the existence of a rotationally supported disk and the steepness of potential holds not only for the early spin-up galaxies, but for our entire sample of 61 MW-mass disk galaxies in TNG50 (see Section~\ref{sec:methods:sample-analysis} and \citetalias{semenov23a} for sample definition). Figure~\ref{fig:rfrac-vc} shows the relation between the rotational support of the young stellar population, the value of $\Rsfr/\Rvmax$, and the halo virial mass for our entire sample across available snapshots. Again, in qualitative agreement with \citet{hopkins23disk}, at low $\Rsfr/\Rvmax$, young stars in galaxies are dispersion dominated, while at high $\Rsfr/\Rvmax$ stars form in a disky, rotation-supported configuration. In TNG50, this transition on average occurs at $\Rsfr/\Rvmax \sim$ 0.4--1, although it is worth noting that the scatter is quite large.

Interestingly, the degree of rotational support also strongly correlates with the mass of the dark matter halo: galaxies become rotation dominated when $\Mvir$ crosses a threshold of $\sim (1\text{--}2)\times 10^{11} \Msun$ (see the right panel in the figure). As we further discuss in Section~\ref{sec:discussion:halo-mass}, this halo mass is consistent with previous numerical works and, importantly, with the estimates of the total mass of the predisk protogalaxy remnant in the MW \citep{conroy22,bk23}. This is also the mass regime mainly explored by \citet{hopkins23disk} in the context of their disk formation model, although they also showed that their conclusions are applicable to ten times higher and lower halo masses.
Note, however, that our results do not show that this halo mass threshold for disk formation is \emph{caused} by the steepening of the potential in the center. As we will show below, a hot circumgalactic halo also forms around the same time, and it can play a role in disk formation (Sections~\ref{sec:icv} and \ref{sec:discussion:disk-formation}).

%-----------------------------------------------------------------
\subsection{Coformation of Stellar Bulge and Disk}
\label{sec:potential:bulge}
%-----------------------------------------------------------------

\begin{figure}
\centering
\includegraphics[width=\columnwidth]{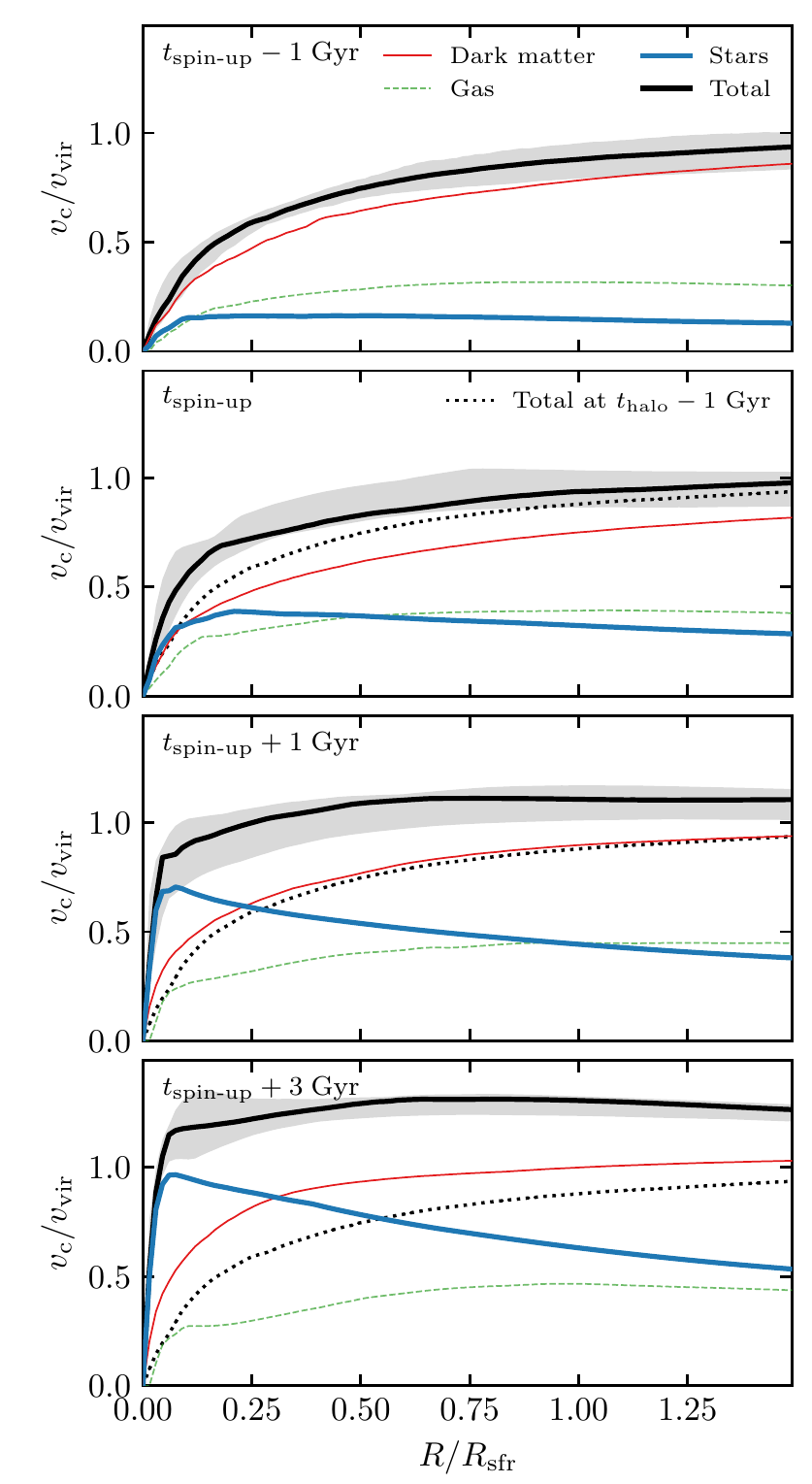}
\caption{\label{fig:vc-profiles} Evolution of rotation curve (proxy for the shape of gravitational potential) before, during, and after galactic disk formation. Rotational velocity and radius are normalized by virial velocity and the size of the star-forming disk, respectively. The panels show the rotation curve, from top to bottom, 1 Gyr before $\tspinup$, at $\tspinup$, and 1 and 3 Gyrs after $\tspinup$. \newtext{The thick black line shows the total $\vc$, while other lines show contributions from dark matter (thin red), gas (dashed green), and stars (thick blue);} to guide the eye, the black dotted line repeats the total rotation curve from the top panel. The rotation curves are stacked across our early spin-up galaxies, with the lines showing the median curves. \newtext{The total $\vc$ and individual contributions exhibit comparable galaxy-to-galaxy variation indicated by the shaded region for the total $\vc$ (16--84 interpercentile range).} Before disk formation, the rotation curve is dominated by the dark matter halo and is gradually rising (the potential is shallow); at later times, the stellar component is quickly growing, making the rotation curve centrally concentrated (the potential becomes steep).}
\end{figure}

\begin{figure*}
\centering
\includegraphics[width=0.3\textwidth]{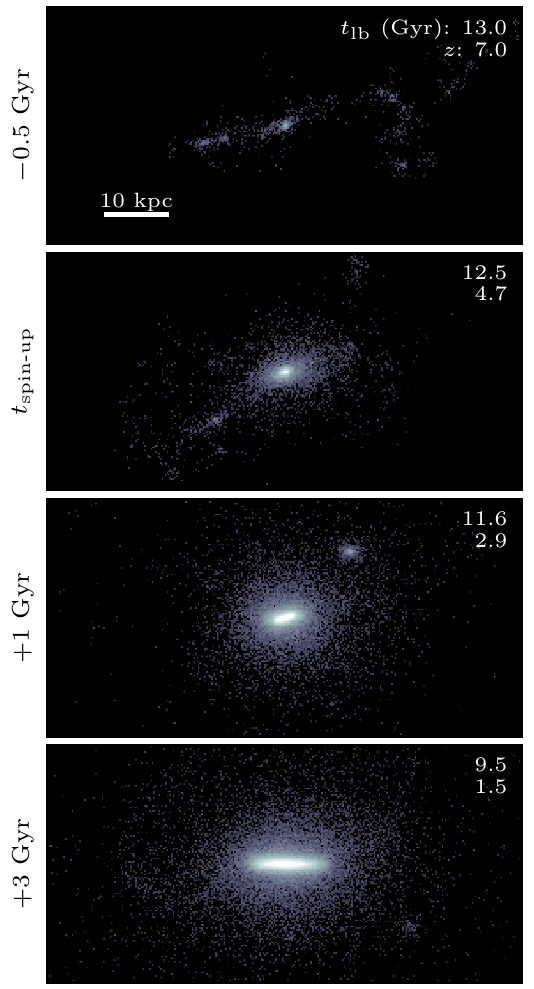}%
\includegraphics[width=0.28\textwidth]{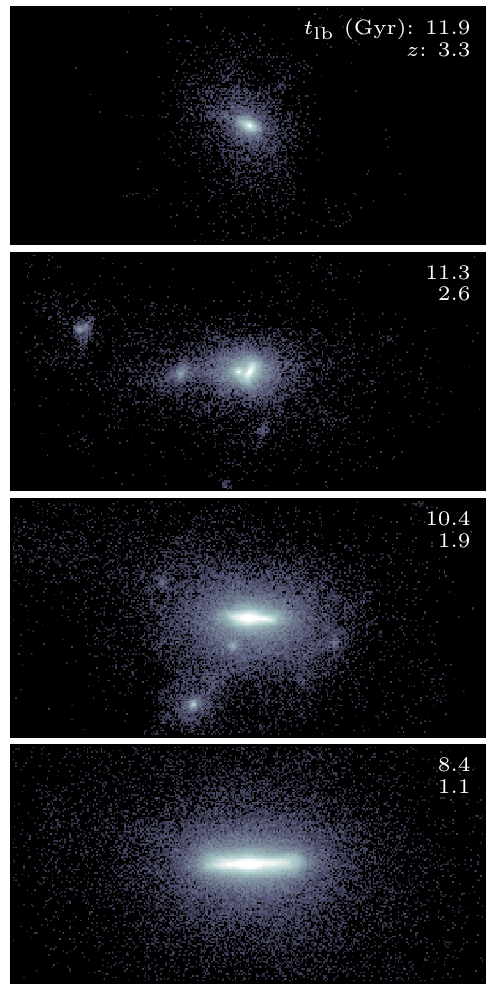}%
\includegraphics[width=0.3\textwidth]{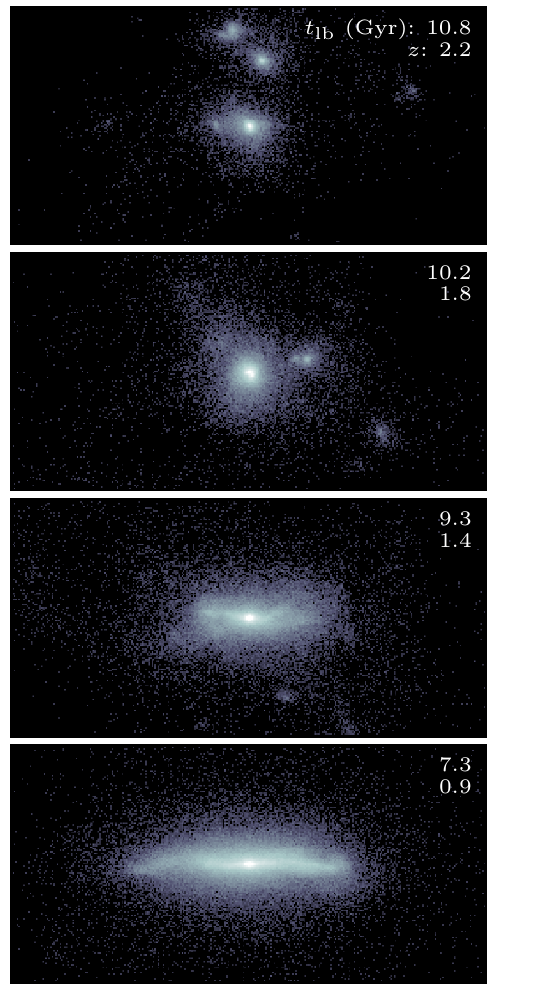}\\
\caption{\label{fig:maps-spinup} Distributions of stars in the three example halos from Figures~\ref{fig:jz} and \ref{fig:jzjc-spinup} at the evolutionary stage shown in Figure~\ref{fig:vc-profiles}. The lookback time and redshift are indicated in the corner of each panel. Prior to and around $\tspinup$, the evolution of protogalaxy is associated with active mergers. At later times, the remnants of this protogalaxy stage form a bulge-like component in the center, around which a stellar disk forms within a few Gyr (see also Figure~9 in \citetalias{semenov23a}). }
\end{figure*}

Overall, Figures~\ref{fig:vc-stack} and \ref{fig:rfrac-vc} indicate that in TNG50 disk formation and a steepening of the central gravitational potential tend to occur together. In Figure~\ref{fig:vc-profiles} we investigate how this steepening proceeds before, during, and after the galactic disk is formed by showing the total rotation curve and the contributions from different components. In addition, Figure~\ref{fig:maps-spinup} shows examples of galaxy morphologies at these different stages of evolution.

Early on, 1 Gyr before $\tspinup$ (top panel), the potential is dominated by the dark matter halo with a gradually raising rotation curve. This is the moment when the disk formation typically sets in; i.e., when the population of young starts acquires the net rotation  (see Figure~\ref{fig:vrot-stacked}). Over the next 4 Gyr of evolution, however, the centrally concentrated contribution from stars rapidly increases and quickly becomes dominant in the galaxy center, making the total rotation curve centrally concentrated as well. The contributions of dark matter and gas evolve approximately self-similarly; i.e., their rotation curves normalized by the virial velocity and galaxy size maintain an approximately constant shape and magnitude. The adiabatic contraction of the dark matter halo---which is manifested in a non-self-similar steepening of dark matter contribution in the very center of the galaxy---becomes significant only at late stages when the stellar component already dominates in the center.

Such an evolution of rotation curves suggests that the steepening of the central gravitational potential occurs as a result of baryonic matter accumulation in the center. As Figure~\ref{fig:maps-spinup} demonstrates and as we also  showed in \citetalias{semenov23a}, the early predisk protogalaxy stage is often associated with active galaxy mergers that result in the formation of a spherical bulge-like component. At later times, a stellar disk forms around this bulge, and a few Gyr after $\tspinup$ a rotationally supported prominent stellar disk is in place.

\begin{figure*}
\centering
\includegraphics[width=0.35\textwidth]{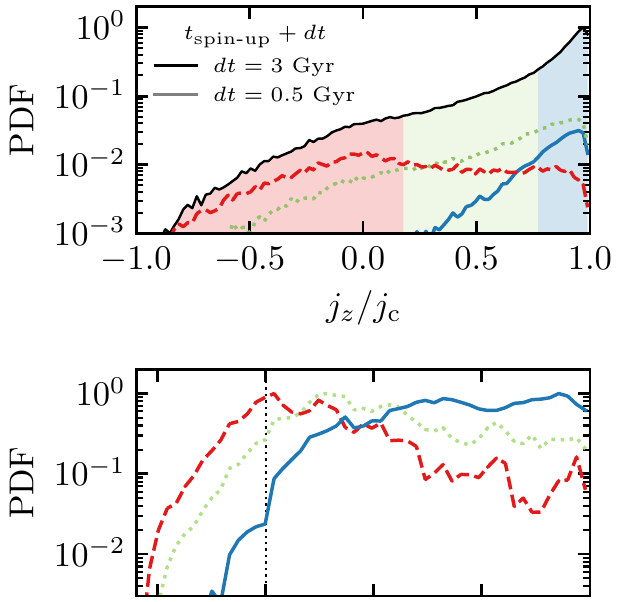}%
\includegraphics[width=0.29124\textwidth]{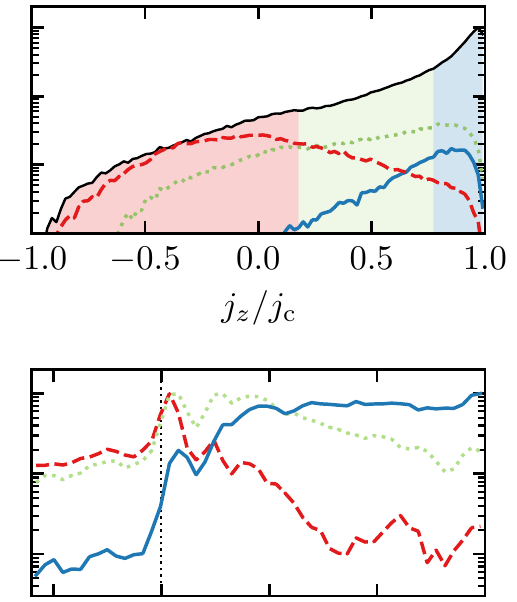}%
\includegraphics[width=0.35\textwidth]{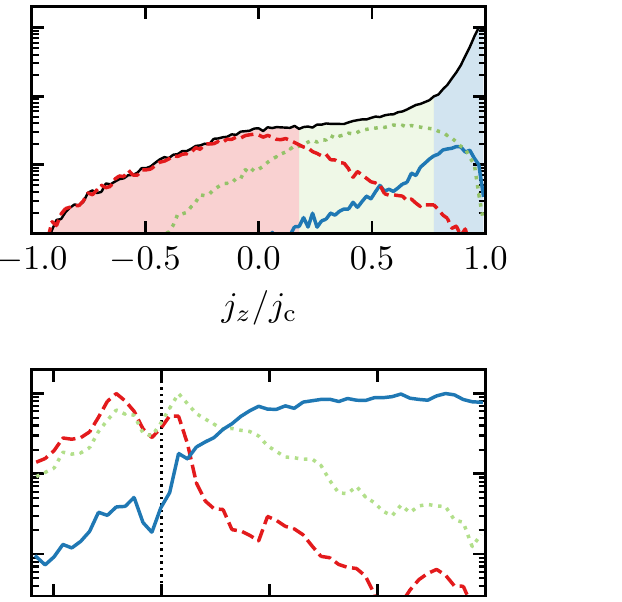}\\
\includegraphics[width=0.35\textwidth]{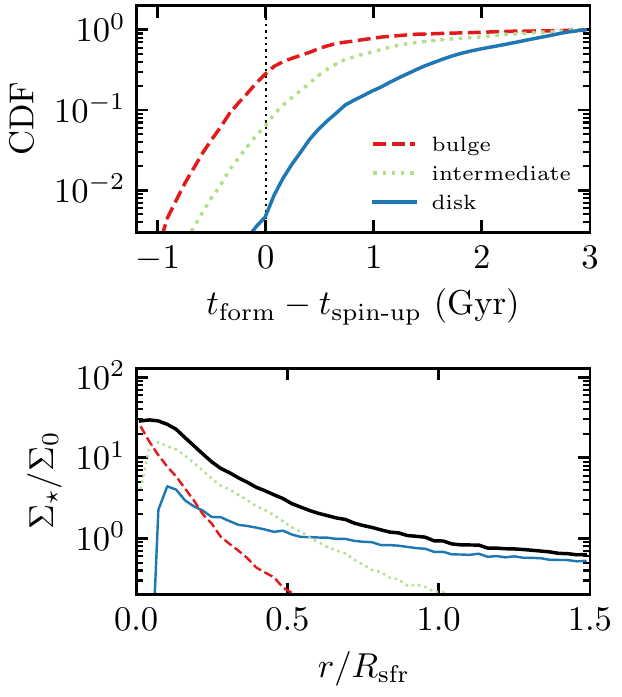}%
\includegraphics[width=0.29124\textwidth]{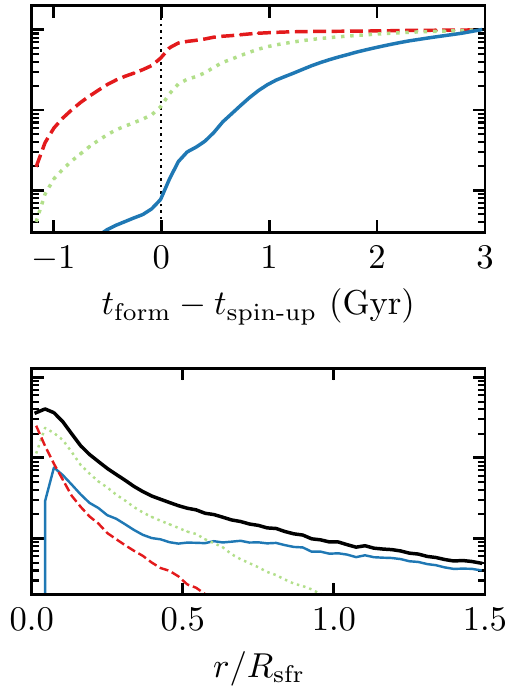}%
\includegraphics[width=0.35\textwidth]{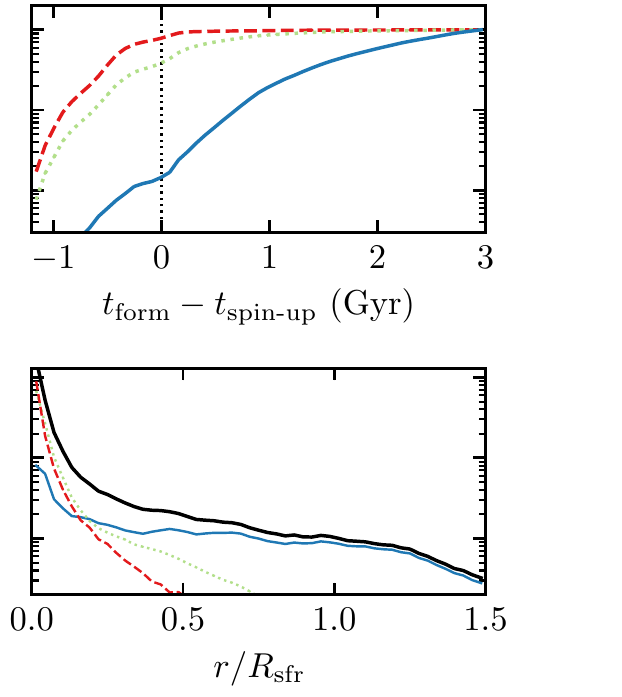}\\
\caption{\label{fig:jzjc-spinup} Coformation of the stellar bulge and disk around $\tspinup$ in our three example halos from Figures~\ref{fig:jz}, \ref{fig:jzjc-spinup}, and \ref{fig:maps-spinup}. {\bf Top:} distributions of stellar orbital circularities, $\jzjc$, when the disk is established, $\tspinup + 3$ Gyr, and shortly after disk formation, $\tspinup + 0.5$ Gyr. At $+ 3$ Gyr, we decompose the stellar population into bulge ($\jzjc < 0.2$; red shaded region), disk ($\jzjc > 0.8$; blue shaded region), and intermediate population (green shaded region). The colored lines show the stars from these three populations at $\tspinup + 0.5$ Gyr; the bulk of the bulge is in place by this point. {\bf Middle:} differential (upper middle) and cumulative (lower middle) distributions of formation times in these three components. All PDFs are normalized by their maximal value. {\bf Bottom:} decomposition of the radial surface density, $\Sigma_\star$, profile into these three components at $\tspinup + 3$ Gyr. The radius is normalized by $\Rsfr$ to facilitate comparison with Figure~\ref{fig:vc-profiles}, while $\Sigma_\star$ is normalized by $\Sigma_0 \equiv M_\star / \pi \Rsfr^2$, where $M_\star$ is the total stellar mass of the galaxy. The figure shows that the bulge forms prior to or around $\tspinup$ and it dominates the mass on the scales where the potential is steep leading to a peaked rotation curve, $r/\Rsfr \lesssim 0.3$ (see the bottom panel of Figure~\ref{fig:vc-profiles}).}
\end{figure*}

This sequence---concurrent bulge and disk formation followed by disk growth---is demonstrated quantitatively in Figure~\ref{fig:jzjc-spinup}. The \newtext{thin black line and colored regions} in the top panels show the distribution of orbital circularities, $\jzjc$, and kinematic decomposition into the bulge \newtext{($\jzjc < 0.2$)}, disk \newtext{($\jzjc > 0.8$)}, and intermediate population of stars\footnote{In this paper, we do not make a distinction between the thick and thin disks as it is not clear whether the resolution of the simulation is sufficient to model thin disks (see also Section~\ref{sec:discussion:disk-formation}). Therefore, to avoid confusion, we label the populations of stars with progressively increasing $\jzjc$ in the top panel of Figure~\ref{fig:jzjc-spinup} as bulge, intermediate, and disk.} based on $\jzjc$ values at a moment after the disk is established, 3 Gyr after $\tspinup$. \newtext{Although the choice of these $\jzjc$ thresholds is somewhat arbitrary, the result remains qualitatively similar when we vary their values.} Colored lines in the same panel show the distribution of $\jzjc$ of stars shortly after disk spin-up, $\tspinup + 0.5$ Gyr. Most of the bulge (red line) was in place at that moment, while most of the disk forms over the subsequent 2.5 Gyr.

The formation sequence is further quantified in the middle row of panels, which shows PDFs and CDFs of formation times for these three populations of stars. The bulk of bulge stars forms prior to or around $\tspinup$, while the disk forms at later times. The inspection of the entire sample of early spin-up galaxies shows that, in many instances, the PDF of formation times of the bulge stars exhibits a prominent peak close to $\tspinup$ as a result of merger-driven enhancement of SFR during protogalaxy coalescence. This feature is particularly clear in the middle column of the figure. The intermediate population follows tracks in-between bulges and disks in part because it mixes the contributions from both populations.

Finally, the bottom panels show stellar surface density profiles scaled by the galaxy size, $\Rsfr$, to facilitate direct comparison with the rotation curves in Figure~\ref{fig:vc-profiles}. These panels demonstrate that at $\tspinup + 3$ Gyr the population of stars identified as bulge indeed occupies the central part of the galaxy causing significant steepening of the surface density profile (see also Figure~9 in \citetalias{semenov23a} for visual impression).  Together with the bottom panel of Figure~\ref{fig:vc-profiles}, this result implies that such bulges dominate the contribution to the total gravitational potential on scales $\lesssim 0.3\; \Rsfr$  and therefore, it is the bulge that provides a steep gravitational potential on these scales, at least at the stage when the disk is in place. 

Note that such a protogalaxy remnant does not necessarily constitute the entire final bulge of the MW and other disk galaxies as bulges can also grow after disk formation, e.g., as a result of star formation in the galaxy center and bar formation driving gas inwards \citep[e.g.,][]{mw-review}. For galaxies that experience late destructive mergers after which their disks reform, the bulge will also consist of the premerger stars. 

To summarize, the results of this section suggest that the formation of a disk and steepening of gravitational potential are correlated. This result is qualitatively consistent with the \citet{hopkins23disk} picture. The steep potential in the center is provided by the stellar bulge formed by the remnant of the predisk protogalaxy and the enhanced SFR in the center during the protogalaxy coalescence. It is however not clear that the steepening itself causes disk formation, as most of the steepening occurs during and shortly after disk spin-up, while, at the onset of disk formation, the potential is still rather shallow and dominated by dark matter. However, given the results of \citet{hopkins23disk}, such  a steepening of the potential can facilitate disk growth during its formation even if the onset of the disk formation itself is not caused by such a steepening.

%-----------------------------------------------------------------
%-----------------------------------------------------------------
\section{Coevolution of Disk and Hot Gaseous Halo}
\label{sec:icv}
%-----------------------------------------------------------------

\begin{figure*}
\includegraphics[width=0.3559\textwidth]{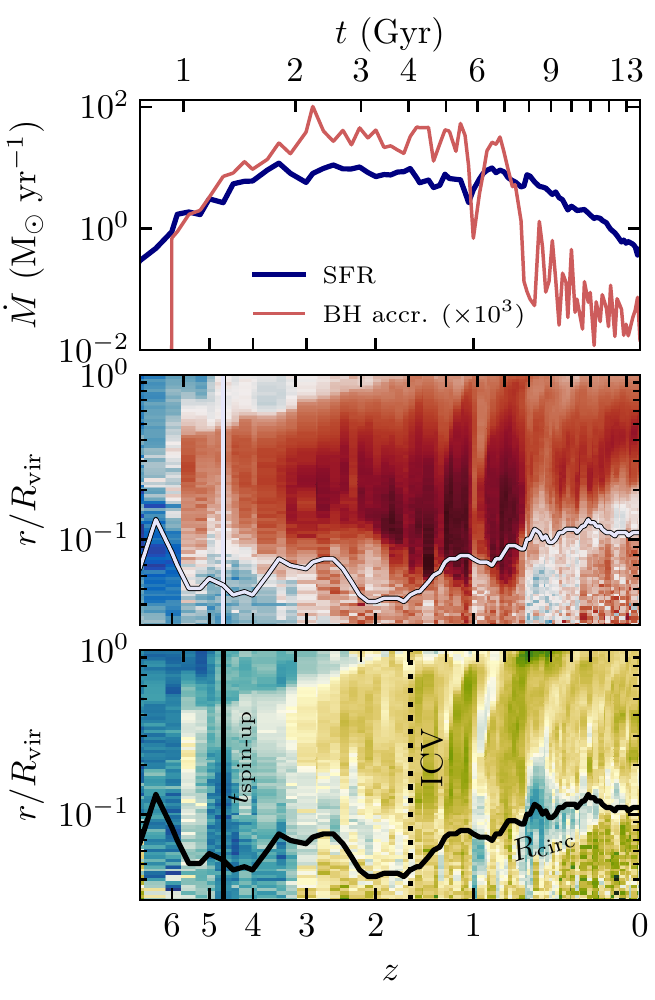}%
\includegraphics[width=0.2880\textwidth]{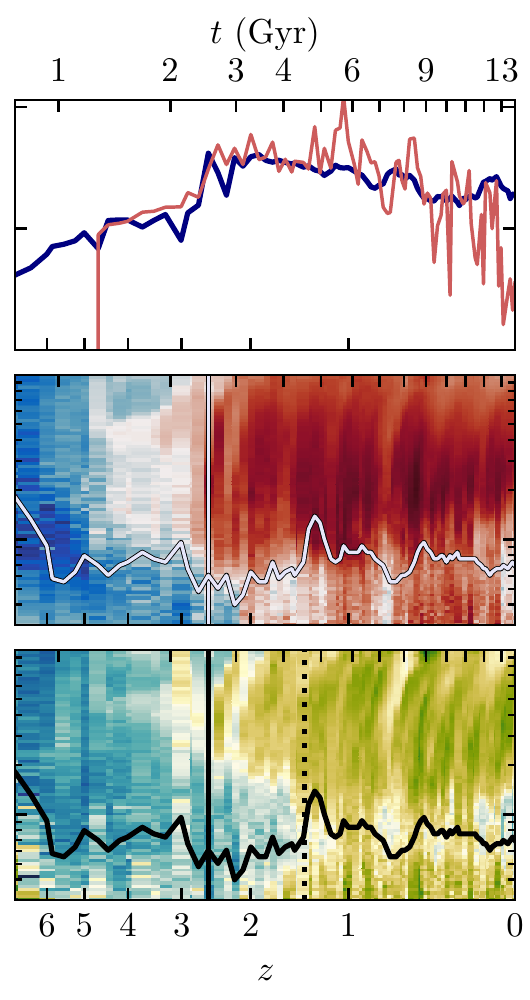}%
\includegraphics[width=0.3559\textwidth]{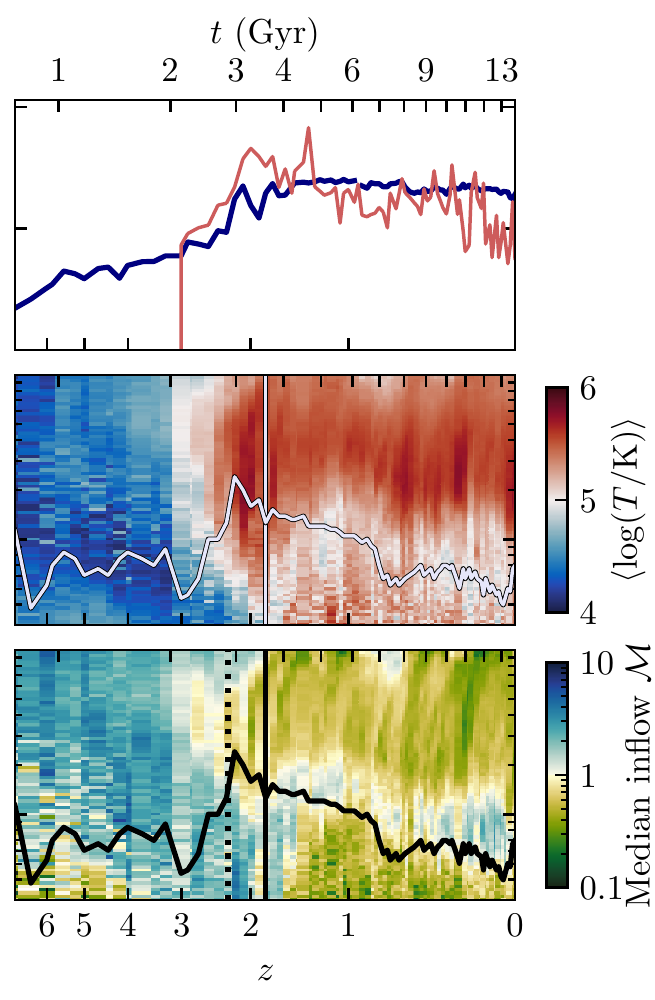}\\
\caption{\label{fig:icv-examples} Examples of formation and evolution of the hot gaseous halo. {\bf Top:} star formation history and the rate of central black hole accretion (as a proxy for AGN activity). {\bf Middle:} radial profile of volume-weighted average log of gas temperature. {\bf Bottom:} volume-weighted median Mach number, $\Mach \equiv v_r/\cs$, of the inflowing gas, i.e., gas with $v_r < 0$. The horizontal curve in the bottom two panels indicates the size of the galaxy (circularization radius, $\Rcirc$; see Section~\ref{sec:methods}). The vertical solid and dotted lines show, respectively, the moment of disk formation, $\tspinup$, and the moment when more than half of the inflowing gas at the galaxy scale becomes subsonic, i.e., the median Mach number at $\Rcirc$ becomes $\approx$ 1. The latter approximately corresponds to the ICV transition in the \citet{stern21} model. The hot gaseous halo with $T \sim \text{few\;}10^5\text{--}10^6$ K forms early, following the increase of SFR and AGN activity. However, in two of these three cases, the inflow of the gas onto the galaxy is still supersonic when the disk is formed, implying that the inflow is not needed to be entirely subsonic for disk formation.}
\end{figure*}

Another model that considers galactic disk formation has been recently developed in the series of works by \citet{stern19,stern20,stern21,stern23} and also tested in zoom-in cosmological galaxy formation simulations \citep[][]{hafen22,gurvich22}. The model posits that a thin disk can form as a result of the inner gaseous halo becoming thermally supported---the concept dubbed in the above works as the ``inner CGM virialization'' (ICV). This transition leads to a qualitative switch of the accretion mode from a rapid supersonic inflow of clumpy material with disoriented angular momentum at early times (cold mode), to a smooth subsonic inflow at late times (hot mode). In the latter regime, the inflowing gas has time to self-align and join the disk smoothly with a narrow distribution of angular momenta, leading to continuous accretion of coherent angular momentum required for the buildup of the disk \citep[][see also Sections~\ref{sec:angmom:alignment} and \ref{sec:angmom:frame} above]{hafen22,stern23}. The exact heating mechanism, be it, e.g., accretion shock or stellar and AGN feedback, is unimportant, as long as it can heat the halo to roughly the virial temperature. This transition is typically predicted to occur rather late, at $z\sim 1\text{--}2$, and the above works argue that it causes the transition from a thick disk with bursty SFR to a thin disk forming stars steadily. 

In this section, we investigate whether a similar picture may explain the spin-up of galactic disks in the first place; i.e., the transition from disordered galaxy formation, often associated with active galaxy mergers (protogalaxy), to a single thick disk. Indeed, the above model predicts that ICV occurs late because it considers a spherically symmetric cooling flow with a single temperature, velocity, and thus Mach number\newtext{, $\Mach \equiv v_r/\cs$,} profile set by the interplay between heating at the accretion shock, adiabatic compression, and radiative cooling. On the other hand, the gaseous halos of simulated galaxies exhibit wide variations in local Mach numbers of the inflowing gas, especially in the presence of stellar and AGN feedback, implying that a fraction of material can be accreted sub- and transonically \newtext{(i.e., with $\Mach \lesssim 1$)} even at early times and may play a role in spinning up galactic disks.

Figure~\ref{fig:icv-examples} shows the evolution of our three representative early spin-up galaxies: SFR and AGN activity (top), halo temperature\footnote{For consistency with \citet{stern21}, we show the volume-weighted average of $\log\,T$.} profile (middle), and the median volume-weighted Mach number of the inflowing gas ($\Mach$; bottom). The Mach numbers are computed individually for each infalling gas cell using the local temperature and radial velocity magnitude.
A comparison of the top two rows shows that, as the SFR becomes sufficiently high ($\SFR \gtrsim 1 \Msunyr$) and an AGN is seeded and becomes active, the halo gas is quickly heated to $T \sim$ few $10^5$--$10^6$ K. Interestingly, the disk starts forming at around the same time, as indicated by the vertical line at $\tspinup$. 

Although the coincidence of hot halo and disk formation might seem to be consistent with the plain ICV picture, in fact, the halo is not hot enough at these early times to make the inflow predominantly subsonic. As the bottom panels show, when disk formation sets in, most of the accretion is dominated by a supersonic inflow out to $\sim 5$ times the galaxy size, at least in the first two out of these three examples.
Within the plain ICV model, such a supersonic inflow should lead to the accretion of clumpy material with disordered angular momentum and provide low-density channels for feedback-driven outflows to propagate efficiently, leading to significant perturbations of the star-forming disk.

The thermalization front---i.e., the lower boundary separating supersonic (blue) and subsonic (yellow) inflow---forms at the halo outskirts when it becomes hot and then propagates outside-in in qualitative agreement with the \citet{stern20} results and opposite to the inside-out propagation of the viral shock\footnote{These results do not necessarily contradict each other. The virial shock determines the outer boundary of the hot halo that, according to Figure~\ref{fig:icv-examples}, indeed moves outwards in agreement with \citet{birnboim-dekel03}. Note, however, that, in the presented halos, this boundary can also be caused by expanding galactic winds and AGN feedback rather than the accretion shock.} \citep[e.g.,][]{birnboim-dekel03}. 
As a result, the thermalization front usually reaches the galaxy after the disk is in place, although sometimes it might precede disk formation. When this front reaches the galaxy, the accretion on galaxy scales becomes predominantly subsonic. This transition is indicated by vertical dotted lines in the bottom panels of Figure~\ref{fig:icv-examples} which show the moment when the median Mach number at $\Rcirc$ (solid black curve) becomes $\approx 1$; labeled as ``ICV'' in the plot.

\begin{figure}
\centering
\includegraphics[width=\columnwidth]{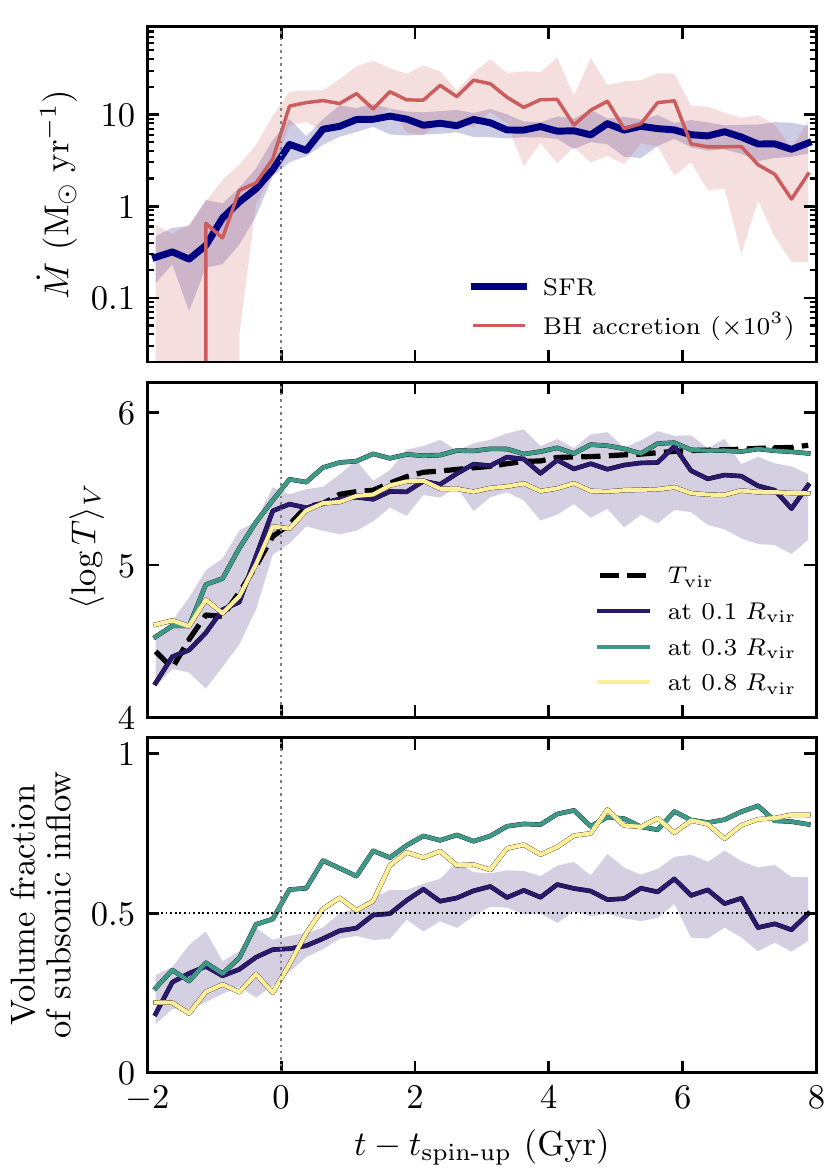}%
\caption{\label{fig:icv-stacked} Analogous to Figure~\ref{fig:icv-examples} but stacked across all 10 of our early spin-up galaxies. Lines show the median tracks, while the bands show the galaxy-to-galaxy variation (16--84 interpercentile range). Average gas temperature (middle panel) and inflow supersonic fraction (bottom) are sampled at 0.1, 0.3, and 0.8\;$\Rvir$. \newtext{Both quantities show comparable galaxy-to-galaxy variation at different radii, and therefore, we only indicate it for 0.1\;$\Rvir$.} The black dashed line in the middle panel shows the median virial temperature, indicating that the combined effect of feedback and adiabatic compression heats the halo gas roughly to virial temperature throughout the evolution. The hot halo forms early, around the time when the disk is formed (vertical dotted line). At this moment, only $\sim 40\%$ of inflowing gas is accreted at subsonic velocities, implying that purely subsonic inflow is not required for disk formation.}
\end{figure}

\begin{figure}
\centering
\includegraphics[width=\columnwidth]{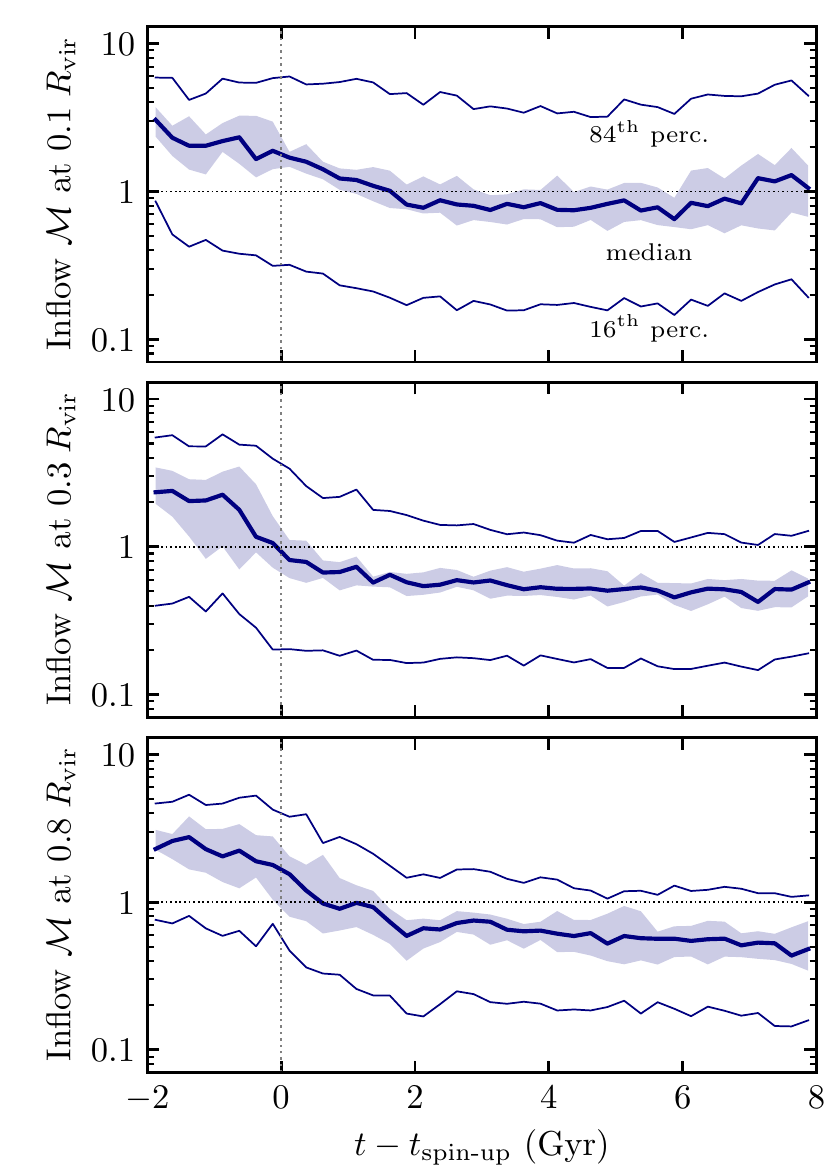}%
\caption{\label{fig:icv-mach-stacked} Distribution of inflow Mach numbers, $\Mach$, at 0.1, 0.3, and 0.8\;$\Rvir$ (from top to bottom). Thick lines show the medians, and thin lines are 16$^\text{th}$ and 84$^\text{th}$ volume-weighted percentiles of $\Mach$ at the given radius. Shaded regions show the galaxy-to-galaxy variation of the median Mach number (16--84 interpercentile range). Mach numbers show wide ranges with the median values before and during disk formation at around $\Mach \sim 2$, implying that a significant fraction of gas is not highly supersonic but accreted in the transonic regime.}
\end{figure}

Figure~\ref{fig:icv-stacked} demonstrates that the above conclusions hold for our entire sample of early spin-up galaxies. The top panel shows stacked evolution tracks of SFR and black hole accretion rates, while the other two panels show the average halo temperature and the relative volume fraction of subsonic inflow material at different radii ranging from close to the galaxy to the outer halo: 0.1, 0.3, and $0.8\,\Rvir$. 
\newtext{The galaxy size (quantified above using $\Rsfr$ and $\Rcirc$) is typically close to $0.1\,\Rvir$ and therefore, this radial slice can be viewed as the transition point between the gaseous halo and disk.}
The halo temperature follows the virial value, $\Tvir \equiv \mu \mp \vc(\Rvir)^2/2\kboltz \approx 6\times10^5\,(\Mvir/10^{12}\Msun)^{2/3} \K$ \citep[e.g.,][]{stern19}, within a factor of 2. Again, although the hot halo forms around the same time as the disk spins up, $\tspinup$, more than half of the inflow occurs at supersonic velocity at that moment.

These results imply that the thick disk formation is not associated with a transition to \emph{dominant} hot-mode accretion. Nevertheless, the volume fraction of the hot-mode subsonic inflow at $\tspinup$ is quite substantial, $\sim 40\%$ at $0.1\;\Rvir$ (see the bottom panel in Figure~\ref{fig:icv-stacked}). Moreover, from the bottom panels of Figure~\ref{fig:icv-examples}, at $\tspinup$, the gas in the inner halo is accreted at only moderate Mach numbers, with median values of $\Mach \sim 2\text{--}5$. This implies that a substantial fraction of the inflow may occur at transonic velocities and therefore might still settle in a disk according to the ICV picture.

The variation of inflow Mach numbers is quantified in Figure~\ref{fig:icv-mach-stacked}. During the onset of disk formation $\sim 1$ Gyr prior to $\tspinup$, the variation of $\Mach$ near the galaxy is indeed quite large (top panel): the median value is $\Mach \sim 2$ and the 16$^\text{th}$ percentile is at $\Mach \sim 0.3$. The distribution of $\Mach$ also shows a clear decreasing trend around the moment of disk formation and settles down $\sim 2$ Gyr after $\tspinup$. The trends at larger radii are qualitatively similar, although the distributions are somewhat narrower. One possible explanation is that star formation and AGN-driven winds may drive stronger variations closer to the galaxy compared to the outer halo.

The results of this section suggest that the spin-up of the initial thick disk does not require the inflow to be predominantly subsonic. However, the formation of the hot halo can still play a role because around half of the material infalls in the hot mode with sub- or transonic velocities. In the context of the ICV picture, such gas may settle into a disk, while the rest of the material infalling in the cold mode will bombard the galaxy from random directions disturbing and thickening the star-forming disk. In addition, at late times, $>2$ Gyr after $\tspinup$, the infall becomes mostly subsonic which might cause the settling down of the disk as argued by \citet{stern19,stern20,stern21}. As Figure~\ref{fig:vrot-stacked} showed, at these times, the scatter of $\vrot$ gradually decreases, which might be a manifestation of such a thick-to-thin disk transition in TNG50.

\vspace{4em}
%-----------------------------------------------------------------
%-----------------------------------------------------------------
\section{Discussion}
\label{sec:discussion}
%-----------------------------------------------------------------

%-----------------------------------------------------------------
\subsection{What Spins Disks Up? Causation versus Correlation}
\label{sec:discussion:disk-formation}
%-----------------------------------------------------------------

In the above sections, we investigated different physical processes that may facilitate the formation of galactic disks in the early universe, in particular, angular momentum feeding from the corotating CGM (Section~\ref{sec:angmom}), changes in the gravitational potential (Section~\ref{sec:potential}), and formation of hot gaseous halos around galaxies (Section~\ref{sec:icv}). With a single simulation without parameter variation, it is inherently hard to establish a causal connection between such processes and the formation of the disk. Nevertheless, exploring these processes in a representative set of MW-like galaxies provides useful insights into how such disks might form in the real universe. 

Apart from providing a representative sample of galaxies, TNG50 also allows us to test these scenarios in simulations with qualitatively different implementations of star formation and feedback physics compared to the FIRE-2 suite, which was used to formulate and initially test the two models connecting galactic disks and potential steepening or hot halo formation \citep{stern21,hafen22,gurvich22,hopkins23disk}. The effective equation of state approach adopted in TNG50 (see Section~\ref{sec:methods:tng50}) leads to significantly less vigorous evolution of early galaxies compared to that driven by bursty star formation in FIRE-2. In this sense, early galaxy formation in TNG50 can be thought of as a steady disk evolution regime opposite to that in the FIRE-2 model, and therefore, TNG50 provides a useful benchmark for comparison with the above studies.

In TNG50, the steepening of gravitational potential and the increase of inflow fraction accreted in the hot mode both occur roughly simultaneously with the formation of the galactic disk. Either or both of these processes might cause the formation of the disk, or they can also be a consequence of disk formation. 
In the subsections below, we outline these different scenarios in more detail and discuss their implications.

%-----------------------------------------------------------------
\subsubsection{Steepening of Gravitational Potential}
\label{sec:discussion:disk-formation:potential}
%-----------------------------------------------------------------

The analysis presented in Section~\ref{sec:potential} was motivated by the work of \citet{hopkins23disk}, who used galaxy simulations from the FIRE-2 suite with varied parameters and showed that galactic disks form as long as the gravitational potential in the region occupied by star-forming gas is sufficiently steep. The authors focused on a halo mass $\Mvir \sim 10^{11} \Msun$, which is the mass regime particularly relevant for disk formation in the early universe (see Figure~\ref{fig:rfrac-vc}), and also showed that their result holds for ten times higher and lower $\Mvir$. 

In TNG50, the formation of a disk is indeed associated with the steepening of the gravitational potential in the center (Figure~\ref{fig:vc-profiles}). This steepening is caused by the buildup of the stellar component which, in the center, is dominated by a bulge formed by the remnant of the predisk protogalaxy and the enhanced star formation in the central region during the coalescence of several major pieces of the protogalaxy (Figure~\ref{fig:jzjc-spinup}).

At the same time, it is not clear whether the steepening of the gravitational potential is what is causing the initial spin-up of the disk. Indeed, when the disk formation sets in (top two panels in Figure~\ref{fig:vc-profiles}), the rotation curve is dominated by the dark matter halo and does not exhibit a prominent peak or flattening, implying that the potential stays relatively shallow. Most of the steepening occurs as the disk is growing, from $\tspinup$ to $\tspinup + 3$ Gyr. Such a steepening might be caused by gas funneling to the center after the formation of an early gas-rich disk.

Thus, our results alone are not sufficient to establish whether disk formation is caused by the steepening of the potential or the other way around. However, together with the findings of \citet{hopkins23disk}, the correlation between potential steepening and disk formation in TNG50 shows that it can be an important factor in disk growth at early times. Even if this steepening is not the main cause of the initial spin-up of the disk and, on the contrary, it is the disk formation that helps to steepen gravitational potential, such steepening can be viewed as a positive feedback mechanism that may speed up the subsequent disk growth.

The above scenario, however, might be challenged by observations of bulgeless disk-dominated galaxies \citep[e.g.,][]{kormendy10,fisher11,kormendy12,simmons13}. In addition, some of the disk galaxies show gradually rising rotation curves without central concentrations and thereby appear to be outliers from the above scenario (see, e.g., the discussion in \citealt{hopkins23disk} and examples of such galaxies in \citealt{bizyaev21}). As was argued by \citet{hopkins23disk}, in galaxies without clear bulges in light profiles, the central mass concentration causing steep potential might be provided by components other than the bulge, such as dark matter, supermassive black holes, gas, and the inner disk among others. Furthermore, the central concentration would not manifest itself in the rotation curves if it were dissolved after disk formation via, e.g., $N$-body relaxation or intermittent energy injection as a result of bursty star formation (a process analogous to the scenario proposed for the formation of cored dark matter profiles; see, e.g., \citealt{pontzen14}), or be hidden under the stars formed at the later disk stage. The investigation of these scenarios and the origin of such galaxies can provide further insights into galactic disk formation, warranting further investigation in cosmological galaxy formation simulations.
It is worth noting, however, that state-of-the-art cosmological simulations appear to be in tension with the observed abundance of thin galaxies \citep{haslbauer22}.

%-----------------------------------------------------------------
\subsubsection{Formation of Hot Gaseous Halo}
\label{sec:discussion:disk-formation:icv}
%-----------------------------------------------------------------

The results of Section~\ref{sec:icv} show that, in TNG50, galactic disks form around the same time as the hot gaseous halos are built up as a result of adiabatic contraction and stellar and AGN feedback. This result is qualitatively consistent with the \citet{stern19,stern20,stern21,stern23} ICV model for the bursty-to-steady disk settling, although, in the context of disk formation, this picture likely becomes more nuanced.

In the plain ICV picture, the gaseous disk settles down as a result of hot gaseous halo formation in which the inflow of gas is predominantly subsonic (hot-mode accretion). Such an inflow brings in material gradually, such that the inflowing gas is able to self-align and continuously and smoothly supply the disk with coherent angular momentum without causing significant perturbations \citep{hafen22,stern23}. In addition, a hot halo around the disk provides pressure confinement that does not allow star formation-driven outflows to propagate freely, preventing cycles of significant removal of ISM gas and perturbations to the disk caused by subsequent reaccretion of this material. In contrast, supersonic cold-mode accretion is associated with the inflow of clumpy material with random angular momentum orientations, causing perturbations to the disk and providing low-density channels for outflows to propagate farther out into the halo and IGM.

In contrast to this plain ICV picture, during disk formation in TNG50, the inflow is not \emph{predominantly} subsonic: more than half of the gas is accreted onto the galaxy in the supersonic cold mode (Figures~\ref{fig:icv-examples} and \ref{fig:icv-stacked}). However, velocity, temperature, and therefore the Mach number of the inflow all exhibit wide variations (Figure~\ref{fig:icv-mach-stacked}), such that a substantial fraction of gas \emph{is} accreted in subsonic hot mode, $\sim$40\% from the bottom panel of Figure~\ref{fig:icv-stacked}. In other words, during disk formation, comparable amounts of gas are accreted in hot and cold mode with most of the gas falling in at transonic velocities.

In the context of the ICV picture, these results imply that the gas accreted in a hot mode may settle down in the halo center and form a coherent disk \citep[as suggested by][]{hafen22,stern23}, while the rest of the cold-mode inflow will continuously bombard the disk with clumpy material with disordered angular momentum causing perturbations to that disk \citep[see also][]{keres05,keres09,dekel-birnboim06,dekel09}. 
At later stages, as the fraction of hot-mode inflow continues increasing and the inflow becomes mostly subsonic (Figure~\ref{fig:icv-mach-stacked} and the bottom panel of Figure~\ref{fig:icv-stacked}), more and more gas is accreted smoothly leading to gradual settling down of the thick to thin disk. 

In this paper, we do not investigate this transition because the resolution of TNG50, despite being high for large-volume cosmological simulations, is likely not sufficient to resolve such thin disks: typical cell sizes in MW progenitor galaxies at $z=1$ \citep[$\sim 80\text{--}100\pc$; see Figure~1 in][]{pillepich19} are comparable to the thickness of the thin disk (few 100 pc). In addition, TNG50 adopts the \citet{sh03} effective equation of state for the ISM of such galaxies, which leads to a smooth pressurized ISM, implying that velocity dispersions of star particles at birth are likely underestimated, resulting in thinner stellar disks \citep[e.g.,][]{bird13,bird21}. Nevertheless, we find a gradual decrease of velocity dispersion of young stars after disk formation that might be a manifestation of galactic disk settling in TNG50 galaxies (see Figure~\ref{fig:vrot-stacked}). This decrease is significantly more gradual than the sharp transition from bursty to steady disk in FIRE-2 simulations \citep{gurvich22,hafen22} suggesting that it is highly sensitive to star formation and feedback modeling.

The connection between the hot halo and disk formation can also be affected by various nonthermal processes. For example, the nonthermal pressure provided by cosmic rays (CRs), which is not included in TNG50, can play an important role. CRs can significantly alter the thermal structure of the gaseous halo and facilitate galactic wind driving  \citep[e.g.,][]{uhlig12,booth13,salem14,butsky18}. As \citet{hafen22} showed, the inflow in CR-dominated halos is analogous to that in a hot halo dominated by thermal pressure. Therefore, if such a CR-dominated halo can be established early, it might transform the formation of galactic disks. In addition, CRs with locally reduced propagation rates can suppress gaseous disk fragmentation resulting in more stable disks with reduced angular momentum transport \citep{semenov21}. Apart from CRs, magnetic fields exhibit strong connections to outflows and directionality in the CGM \citep[e.g.,][]{ramesh23b,ramesh23a} and therefore might also play a role in disk formation. 

Finally, our results also motivate considering galactic outflows in models connecting disk formation and settling to the properties of the gaseous halo. Our results suggest that outflows can facilitate the exchange of the angular momentum between the inflow and the disk (Sections~\ref{sec:angmom} and \ref{sec:discussion:winds}). In the context of the \citet{stern20} model, such an effect can relax the requirement that the inflow must be subsonic down to the disk scale; instead, it only needs to be subsonic down to the region occupied by the recycled fountain outflows, as on smaller scales the inflowing angular momentum is already aligned with the disk.
This and the role of partial hot- and cold-mode accretion in disk formation and subsequent settling of the disk as hot mode becomes dominant can be further investigated in zoom-in resimulations of these galaxies with more detailed modeling of ISM physics, which we leave to future work.

%-----------------------------------------------------------------
\subsection{Importance of Outflows: Feeding of Angular Momentum from Corotating CGM}
\label{sec:discussion:winds}
%-----------------------------------------------------------------

\begin{figure}
\centering
\includegraphics[width=\columnwidth]{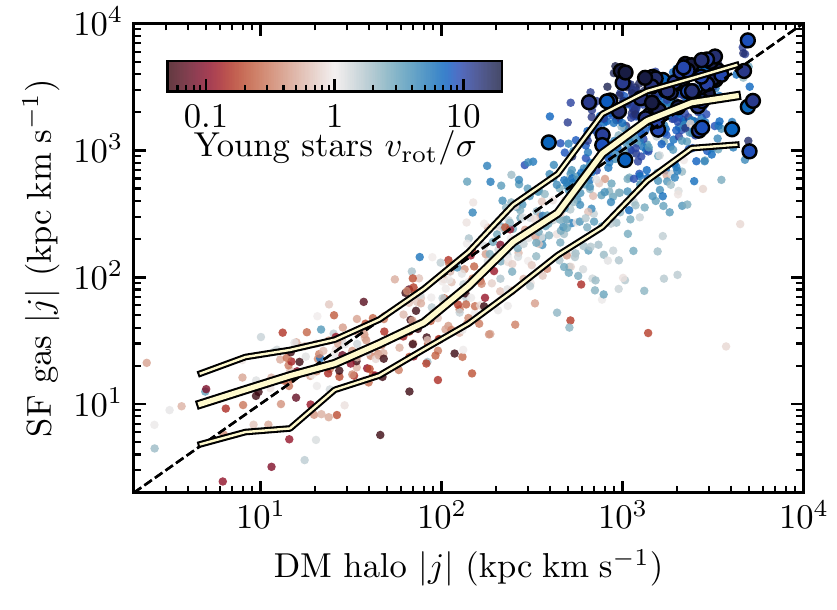}
\caption{\label{fig:jdm-jsf} The relation between the total specific angular momentum contained in the star-forming gas and in the dark matter halo. The points show all 61 of our MW-mass galaxies identified in TNG50 across available snapshots spanning the range of redshifts from $z \approx 12$ to $0$. The final locations of the galaxies are marked with \newtext{large circles} while the color of other points indicates the degree of rotational support of the young ($<$100 Myr old) stellar disk, $\vrot/\sigma$. \newtext{Yellow} lines show the running median and 16--84 interpercentile range, while the dashed line shows the one-to-one relation. Although on average the angular momentum of star-forming gas follows that of dark matter, there is a significant scatter of $\sim0.3\text{--}0.5$ dex before and after disk formation caused by various baryonic processes. }
\end{figure}

During disk formation and subsequent evolution, galactic outflows play an important role in the buildup of galactic angular momentum.
In fact, galactic outflows are required to produce disk galaxies in cosmological simulations: without outflows, simulations suffer from catastrophic angular momentum losses and lead to the formation of overly massive, compact, and centrally concentrated galaxies \citep[e.g.,][]{navarro-white94,navarro95,navarro-steinmetz00}. Galactic outflows alleviate this problem as they prevent strong fragmentation and keep a significant fraction of gas in the halo resulting in better retention of angular momentum \citep[e.g.,][]{governato04,governato07,okamoto05,scannapieco08,zavala08}. In addition, outflows can remove low-angular momentum gas and keep the disk contribution to the gravitational potential sufficiently low, suppressing the disk instabilities that would otherwise transport angular momentum, causing the formation of massive spheroids \citep[e.g.,][]{brook12,agertz-kravtsov16}.

The connection between the outflows and angular momentum of galaxies has been systematically explored using a statistical sample of $M_\star \sim 10^9\text{--}10^{12}\Msun$ galaxies from the Illustris simulations by \citet{genel15}. These authors showed that stronger star formation-driven winds help to retain a higher fraction of the angular momentum generated during large-scale structure formation. In follow-up work, \citet{defelippis17} demonstrated that the angular momentum in simulations with galactic winds is higher largely because recycled outflows regain some of the angular momentum that was lost during the initial infall. A similar effect was also found in zoom-in cosmological simulations by \citet{ubler14} and \citet{christensen16} that showed that ejected low-$j$ gas can accumulate significant angular momentum before reaccretion via torquing and angular momentum exchange with the high-$j$ gas in the outer halo.

Figure~\ref{fig:jdm-jsf} shows the relation between the specific angular momentum of star-forming gas and that of the host halos of all of our MW-like TNG50 galaxies in all available snapshots. In qualitative agreement with previous studies, on average, the angular momenta of both components are similar, with a scatter of $\sim0.3\text{--}0.5$ dex around the one-to-one relation, both before and after disk formation (red and blue points, respectively).

The results of Section~\ref{sec:angmom} and, in particular, the close similarity of the distributions of angular momenta carried by inflows and outflows, are qualitatively consistent with the above picture, suggesting the efficient exchange of angular momentum between the inflows and outflows. This exchange can happen as a result of either mutual torques or direct mass exchange. In the latter case, a significant part of the inflow can result from the recycling of previous outflows, and vice versa, outflows may stop and turn around some of the infalling gas. 

This picture implies that outflows can facilitate the angular momentum exchange between the inflowing gas and the disk.
Indeed, the outflows move ISM gas and its angular momentum back to the gaseous halo and IGM, creating a corotating CGM around the galaxy, which may affect galaxy formation and evolution in several ways. First, recycling of such fountain outflows will feed the galaxy with prealigned angular momentum inherited from the disk. Second, the gas inflowing from the IGM must pass through the entire corotating CGM, which will act as an intermediary in the angular momentum exchange with the galaxy. Third, such a corotating CGM can work as a reservoir from which fountain outflows may acquire angular momentum before recycling.
Quasar line absorption observations indeed show evidence for corotation of warm, $T \sim 10^{4}$ K, CGM gas with their parent galaxies \citep[e.g.,][]{martin19,zabl19,lopez20}.

This picture of corotating outflow recycling as a possible scenario for disk formation and growth is tightly linked to the geometry of the gas flows as outflows and inflows tend to occur along different directions. Outflows tend to vent perpendicular to the disk while accretion preferentially occurs in the disk plane \citep[e.g.,][]{nelson19,peroux20}. However, again, the causal connection between the formation of the corotating CGM and a disk is hard to clearly establish. Indeed, the outflow can become collimated \emph{as a result} of disk formation, leading to the buildup of corotating CGM. In addition, corotating CGM can be also a result of coherent angular momentum accretion from the IGM \citep{defelippis20}.

Note also that, although outflow recycling can explain why the total amounts and the distributions of angular momentum contained in inflowing and outflowing gas are similar throughout the halo, this similarity could be coincidental. For example, these distributions might be shaped solely by the adiabatic expansion of outflows and compression of inflows. Further analysis, e.g., using passive tracer particles (which are available in TNG) along the lines of work by \citet{hafen22}, could provide useful insights into this process. We leave such an analysis to a future study.

%-----------------------------------------------------------------
\subsection{Halo Mass Threshold for Disk Formation}
\label{sec:discussion:halo-mass}
%-----------------------------------------------------------------

As Figure~\ref{fig:rfrac-vc} showed, disks in our full sample of 61 MW-mass TNG50 galaxies tend to form when their host halos become more massive than $\Mvir \gtrsim (1\text{--}2) \times 10^{11} \Msun$. Such a halo mass is consistent with the previous TNG50 results of \citet{pillepich19} that galaxy morphologies become dominated by disks when their stellar mass exceeds $M_\star \gtrsim 10^9 \Msun$, which corresponds to the above halo mass scale \citep[e.g.,][]{behroozi19}.
Interestingly, this threshold is also consistent with the findings from zoom-in cosmological simulations with substantially different implementations of star formation and feedback \citep[e.g.,][]{el-badry18,dekel20}, suggesting a generic origin of such a threshold.

Importantly, this mass is in the ballpark of the total stellar mass estimate of the Aurora population (i.e., the predisk protogalaxy remnant) by \citet{bk23} and higher than, yet reasonably close to the estimate from \citet{conroy22}: although the stellar mass reported by these authors is considerably lower, $\sim 6 \times 10^7 \Msun$, the stellar mass--halo mass relation at these masses is steep, and the corresponding average halo mass, $\sim 5 \times 10^{10} \Msun$ \citep{behroozi19}, is within a factor of a few from the value that we find.

The existence of such a threshold also explains why ALMA and JWST observations of disk galaxies at redshifts $z > 4$ do not contradict significantly later formation of typical MW-mass disks (see \citetalias{semenov23a} and \citealt{bk22}). Indeed, such galaxies are more massive than typical MW progenitors at that epoch, implying that they become massive enough to form a disk earlier than the MW. These galaxies are likely progenitors of present-day massive ellipticals in the centers of galaxy groups and clusters. Galaxy formation simulations do form disks at such high redshifts
\citep[e.g.,][]{kohandel20,kretschmer22,tamfal22}. For example, \citet{feng15} showed that, in the large-volume cosmological simulation BlueTides, $\sim$70\% of galaxies more massive than $M_\star > 10^{10} \Msun$ at redshifts $z \sim 8\text{--}10$ are disks, which is consistent with the above halo mass threshold.

Although in Section~\ref{sec:potential} we considered this mass threshold in the context of stellar bulge formation causing steepening of the gravitational potential, our results should not be interpreted as proof of the causal connection between the two. For example, this mass threshold might also reflect the mass at which a hot gaseous halo forms (see Section~\ref{sec:discussion:disk-formation}). 
Such a halo mass threshold is also consistent with the predictions of \citet{dekel20} who argued that galactic disks tend to survive only in halos with $\Mvir > 2\times 10^{11} \Msun$ as a result of the reduction in both the efficacy of stellar feedback in perturbing the disk and the rate of mergers causing galactic spin flips. While the feedback prescription in TNG50 might not be suitable to investigate early disk disruption, Figure~\ref{fig:alignment-examples} above shows that the orientation of galactic angular momentum indeed tends to change frequently prior to disk formation.

%-----------------------------------------------------------------
%-----------------------------------------------------------------
\section{Summary and Conclusions}
\label{sec:summary}
%-----------------------------------------------------------------

Using a representative sample of MW-like galaxies from the TNG50 cosmological-volume simulation \citep{pillepich19,nelson19}, we investigated galactic disk formation in the early universe. We focused on a subsample of 10 direct MW analogs in terms of their stellar chemo-kinematic distributions, suggesting the early formation of the stellar disk in agreement with the recent local galactic archeology data---galaxies that we dubbed ``early spin-up'' in \citetalias{semenov23a} (see Figures~\ref{fig:sample} and \ref{fig:FeH-vrot} above). We explored various physical processes operating during disk formation in these galaxies motivated by recent theoretical models. Our key findings can be summarized as follows.

\begin{enumerate}
    \item The specific angular momenta of the inflow, star-forming gas, and outflow are all rapidly growing during the disk formation stage and then settle $\sim 2$ Gyr after the disk is formed. During disk formation, the inflow and outflow carry $\sim 2\text{--}3$ times higher angular momentum than that contained in the star-forming gas, while, at later stages, the angular momenta of all three components become comparable, ending the rapid growth phase (Figure~\ref{fig:vrot-stacked} and the top three panels of Figure~\ref{fig:jz-stacked}).
    
    \item Apart from a similar magnitude of total specific angular momentum, the inflowing and outflowing gas show remarkably similar distributions of the angular momentum orientation with respect to the disk, suggesting an efficient exchange of angular momentum and/or mass between the inflow and outflow, e.g., via outflow recycling (Figure~\ref{fig:jz-stacked}, bottom panel, and Figure~\ref{fig:jz}). As the outflows inherit angular momentum from the disk, they can play an important role in facilitating the angular momentum exchange between the inflow from the IGM and the disk.
    
    \item The angular momentum of both inflow and outflow are on average misaligned with respect to the galaxy prior to disk formation but become highly aligned shortly after the disk is formed (Figure~\ref{fig:jz}). This transition is due to both the disordered orientation of the inflowing material and random variations in the reference orientation set by the angular momentum of the galaxy before a well-defined disk is established (Figure~\ref{fig:alignment-examples}). After disk formation, the distributions of alignment angles narrow down closer to the galaxy, with significant alignment occurring rather far away from the disk, $\sim3\text{--}5$ times larger scales than the galaxy size (the bottom panels of Figure~\ref{fig:jz}). 

    \item The disk formation in TNG50 is correlated with the steepening of the gravitational potential within the region occupied by the galaxy, in qualitative agreement with the \citet{hopkins23disk} results (Figures~\ref{fig:vc-stack}--\ref{fig:rfrac-vc}). The potential steepening is caused by the formation of a centrally concentrated stellar distribution that forms around the time of disk formation (Figure~\ref{fig:vc-profiles}). The stellar mass at the scales where this steepening occurs is dominated by a bulge that is formed from the remnant of the predisk protogalaxy and/or enhanced star formation in the galaxy center during the coalescence of different pieces of the protogalaxy (Figure~\ref{fig:jzjc-spinup}).

    \item The disk formation in TNG50 is also correlated with the formation of a hot gaseous halo with an approximately virial temperature of a few $10^5$ K. Despite the halo being hot, during disk formation, more than half (by volume) of inflowing gas is accreted in supersonic cold-mode (Figures~\ref{fig:icv-examples} and \ref{fig:icv-stacked}). Nevertheless, the local Mach number of the inflowing gas exhibits a large scatter leading to a significant fraction of gas (almost a half) being accreted in sub- and transonic hot mode \newtext{with corresponding Mach numbers of $\lesssim 2$} (Figure~\ref{fig:icv-stacked}, bottom panel, and Figure~\ref{fig:icv-mach-stacked}). In the context of the theoretical model positing thin disk formation as a result of cold-to-hot-mode transition \citep{stern19,stern20,stern21,stern23}, this result implies that the hot-mode part of the inflow can settle into a disk, while the rest of the cold-mode inflow might continue perturbing and thickening the disk via accretion of clumpy material with chaotic angular momentum.

    \item Galactic disks tend to form when their host halos become more massive than $\Mvir \sim (1\text{--}2) \times 10^{11} \Msun$ (Figure~\ref{fig:rfrac-vc}), in agreement with previous findings in TNG50 \citep{pillepich19} and zoom-in cosmological simulations \citep[e.g.,][]{el-badry18,dekel20}. Importantly, such a threshold is in the ballpark of the stellar mass estimate of the predisk protogalaxy remnant in the MW from \citet{bk23} and higher than that from \citet{conroy22}, although the latter is reasonably close in terms of the halo mass (Section~\ref{sec:discussion:halo-mass}).
\end{enumerate}

Finally, we emphasize that our results should not be considered as evidence that either corotating outflow recycling, potential steepening, or hot halo formation  \emph{causes} disk formation (see also Section~\ref{sec:discussion:disk-formation}). It is inherently hard to establish such a causal connection using a single simulation without parameter variation. However, thanks to a representative sample of galaxies, our results demonstrate that these processes are occurring at the same time as the disk forms and can play important roles even if they do not spark disk formation individually. Future studies using zoom-in resimulations of these galaxies with physics variations could help to pinpoint the causal connections and gain further insights into galactic disk formation.

\section*{Acknowledgements}
We thank the anonymous referee for their constructive feedback that improved the manuscript.
We are grateful to Vasily Belokurov, Andrey Kravtsov, and Jonathan Stern for their comprehensive comments on the draft that improved this paper.
We also thank Greg Bryan, Claude-Andr\'e Faucher-Gigu\`ere, Drummond Fielding, Shy Genel, Alexander Gurvich, Zachary Hafen, Philip Hopkins, Andrew Wetzel, and the participants of the Disk Formation Workshop at UC Irvine for inspirational and insightful discussions.
The analyses presented in this paper were performed on the FASRC Cannon cluster supported by the FAS Division of Science Research Computing Group at Harvard University and on the HPC system Vera at the Max Planck Computing and Data Facility, and we thank Annalisa Pillepich for providing access to Vera.
Support for V.S. was provided by NASA through the NASA Hubble Fellowship grant HST-HF2-51445.001-A awarded by the Space Telescope Science Institute, which is operated by the Association of Universities for Research in Astronomy, Inc., for NASA, under contract NAS5-26555, and by Harvard University through the Institute for Theory and Computation Fellowship. 
D.N. acknowledges funding from the Deutsche Forschungsgemeinschaft (DFG) through an Emmy Noether Research Group (grant number NE 2441/1-1).
The analyses presented in this paper were greatly aided by the following free software packages: {\tt NumPy} \citep{numpy_ndarray}, {\tt SciPy} \citep{scipy}, and {\tt Matplotlib} \citep{matplotlib}. We have also used the Astrophysics Data Service (\href{http://adsabs.harvard.edu/abstract_service.html}{\tt ADS}) and \href{https://arxiv.org}{\tt arXiv} preprint repository extensively during this project and the writing of the paper.

\bibliographystyle{aasjournal}
\bibliography{}

\end{document}